%% file: main.tex
\documentclass{ecai}  %

\usepackage{graphicx}
\usepackage{latexsym}

\paperid{2}        %

\usepackage{algorithm}
\usepackage{algorithmic}

\usepackage{newfloat}
\usepackage{listings}

\usepackage{booktabs}
\usepackage{tabulary}

\usepackage{adjustbox}

\usepackage{pgfplots}
\DeclareUnicodeCharacter{2212}{−}
\usepgfplotslibrary{groupplots,dateplot}
\usetikzlibrary{patterns,shapes.arrows}
\pgfplotsset{compat=newest}

\usepackage{eqparbox,array}

\usepackage{tabularx}

\usepackage{cleveref}

\address[A]{Booz Allen Hamilton}
\address[B]{University of Maryland, Baltimore County}
\address[C]{Syracuse University}

\begin{document}

\begin{frontmatter}
\title{Does Starting Deep Learning Homework Earlier Improve Grades?}
\author[A,B,C]{\fnms{Edward}~\snm{Raff}} %
\author[B]{\fnms{Cynthia}~\snm{Matuszek}} %

\begin{abstract}
Intuitively, students who start a homework assignment earlier and spend more time on it should receive better grades on the assignment. However, existing literature on the impact of time spent on homework is not clear-cut and comes mostly from K-12 education. It is not clear that these prior studies can inform coursework in deep learning due to differences in demographics, as well as the computational time needed for assignments to be completed. We study this problem in a post-hoc study of three semesters of a deep learning course at the University of Maryland, Baltimore County (UMBC), and develop a hierarchical Bayesian model to help make principled conclusions about the impact on student success given an approximate measure of the total time spent on the homework, and how early they submitted the assignment. Our results show that both submitting early and spending more time positively relate with final grade. Surprisingly, the value of an additional day of work is apparently equal across students, even when some require less total time to complete an assignment. 
\end{abstract}
\end{frontmatter}

\section{Introduction}

In developing a course on deep learning for the University of Maryland, Baltimore County (UMBC), we focused on practical coding experience and implementation of deep learning methods for the course content and evaluation. Compared to assignments in some machine learning classes, the course requirement to use a Graphics Processing Unit (GPU) led us to strongly emphasize throughout the semester that students should start their homework early, as they need to have sufficient time to run their code, iterate and try to fix bugs if errors occurred, and ask the instructor for assistance. As the course progressed and assignments were due, students would sometimes ask how early should they start an assignment, and we had no quantifiable justification for our answers. This paper remedies this issue, and studies the overall question: does starting and/or submitting an assignment earlier improve student's grades on that homework?

The literature studying the impact of time spent on homework, at large, is sparse. One set of work studies the impact of ``procrastination,'' measured by comparing the time an assignment is due and the time the assignment was submitted \cite{Jones2019,Jones2020YearTE,Cormack2020}. This is the easiest form of data to study as it is readily available in modern electronic submission systems. The current studies regularly conclude that those who submit earlier obtain better grades. While we collect the same data, we do not study it directly as it is a proxy for time spent. That is to say, a student who does the assignment the day before and submits the day before has procrastinated, but would not show up in the data as a procrastinator. Similarly a student who starts weeks in advance, and submits at the literal ``11'th hour'' did not procrastinate, but would be marked a procrastinator when using only submission time.  In our study we have the start time of an assignment, and we use that with the time submitted to compute a ``total time'' spent on the assignment. This is not a contiguous measure of time or effort, but we argue a likely better measure of the quantity we care about: total effort spent on an assignment.

Others have also attempted to look at the total time spent on homework and its relation to performance, and have regularly concluded that too much time spent on homework can result in \textit{reduced} scholastic performance~\cite{Galloway2013,Ozyildirim2021,Fernandez-Alonso2015}. All of these works focus on students in high school or earlier, and are focused on overall scholastic outcomes rather than per-assignment results. Similarly, the data is the result of survey information, where our total time is determined via the edit history of the assignment. For this reason we believe our total time measure to be a more reliable, though still not perfect, measure of the goal. In a larger sense, there are numerous differences between our population and those studied before (we study graduate students vs. K-12, homework grade vs. overall performance, and coding and deep learning vs. general Science Technology Engineering and Math subjects). The consistency of the prior studies results' about negative returns for ``over-studying'' necessitate exploration of the question.%

The rest of our paper is organized as follows. First we will give extensive background on our data, the course, and necessary background to interpret and understand the results in \cref{sec:data}. We have $N=68$ total subjects over three semesters of the course. Next we detail the model we use for understanding the data in \cref{sec:model}, which uses a hierarchical Bayesian approach, as is generally encouraged in studies of this nature~\cite{Flunger2021}. Our results will be presented in \cref{sec:results}, where we conclude that more time spent is better than less, and submitting earlier and spending more time have a statistically significant positive impact on a student's grades. Finally we will review other related works in \cref{sec:related_work} and then conclude in \cref{sec:conclusion}.

\section{Data Collection \& Background} \label{sec:data}

To study the question, our data is collected from a course taught by the author(s) in the Spring 2020, Fall 2020, and Spring 2021 semesters at
UMBC. The course content and questions were developed into a book Inside Deep Learning\footnote{Available at \url{https://www.manning.com/books/inside-deep-learning}} \cite{IDL}.
We note that this immediately introduces a set of biases into our results. Most notably, the semesters involved have occurred at the onset of and through the COVID-19 pandemic. This has introduced stresses on students and faculty that are beyond the scope of this study. In addition, one set of instructor(s) are involved, and so any instructor modulated response will not be observable.

As part of the course design, students were instructed to write, test, and submit all of their homework within a Google Colaboratory environment. This choice was originally made as a mechanism to satisfy the desiderata:

\begin{enumerate}
    \item Free or cheap GPU availability to students
    \item Avoiding versioning conflicts and software installation issues
    \item Having a simple means of running student assignments (also avoiding student vs. teacher package mismatches)
    \item Provides an easy way for students to get help / feedback on assignments
\end{enumerate}

An unintended benefit of this course design choice was that Colab kept a \textit{sparse} edit history of the assignments. Different from a normal version control system, Colab will take snapshots that can be differentiated against each other (or the current version) of a document at regular intervals, or as an explicit save request is called. The exact mechanics of this process are not documented, but there does appear to be an age-off process where some subset of snapshots are removed over time, eventually resulting in no edit history. 

This edit history, collected close in time to course completion, was initially used as a means of assisting and helping students with feedback on how to perform better in the course. For example, we could see when a student started the homework assignment 40 minutes before the assignment was due, and thus, was unable to complete the assignment. In such cases the student was coached and advised on time management and the need to reach out to instructors early if something may prevent them from timely completion. Retroactively, this also became the driving force for this study: is there a significant difference in student's grades based on when they begin their assignment?

To answer the question, we went through every student's homework assignments version history, which is available when students submit an editable link to their Colab files (as required by the course submission). It was quickly determined that the first edit in the history was not a reliable method of determining the student's start date, as many students would create the homework file days or weeks before starting the assignment. Small edits to copy questions would also occur. Other faux starts included copying code from the course book that would be used by the homework solution (e.g., assignment says to modify the code), but had not yet started actual modification.  For this reason, we subjectively reviewed all edit histories until the first edit that appears to show the student trying to make progress on the assignment, and recorded that as the start date of the assignment. The submission time on the assignment was obtained via email or Blackboard, depending on which method was used to receive homeworks during the given semester. 

Combined, this gives us the time of submission and the time between start and submission. A limitation we make explicit is that we cannot reliably measure or quantify the true time spent working on the assignment, because the Colab history is coalesced and undocumented in its triggering frequency/characteristics. There may be instances where a student in our data ``starts early,'' but does not work on the assignment for an extended period of time. Since we have no way to detect this, we leave the issue to future work. 

\begin{figure}[!h]
    \includegraphics[width=\columnwidth]{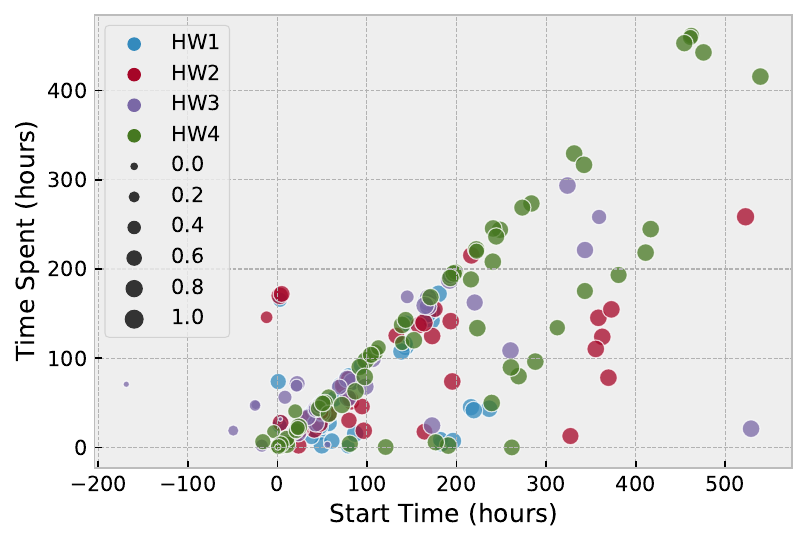}
    \caption{Visualization of the raw data used in this study. The number of hours in advance of the due date that a student started is on the x-axis, and the number of hours the student appears to have worked on the assignment is the y-axis. Color denotes which homework assignment, and size denotes the grade received on that assignment (1.0 = 100\%, 0.0 = 0\%). Note most students did well, so many large circles are present. Negative x-axis values occur when students start the assignment after it was due. }
    \label{fig:data_view}
\end{figure}

A visualization of the resulting data is given in \cref{fig:data_view}. Note that negative values on the x-axis indicate use of the late submission policy. Most students submit near the deadline, resulting in a strong linear trend. Some students realize they have completed the homework early, and fall to the bottom-right quadrant of the plot. 

\subsection{Homework Details}

Each semester four homeworks were assigned, and their grades were the basis of this study. The design choices of these homeworks impact our modeling of the problem, and the reader's ability to subjectively interpret the applicability of the results to their own curricula and assignments. All four assignments were designed such that someone who knew how to perform all tasks could complete them within 40 minutes. This reflects the time of an expert practitioner/researcher who has previous done each assignment in the context of a job or research goal, and has over a decade of experience writing code and in machine learning. As such the 40 minutes is usually not reflective of how long a student will take to complete the assignment, but is done to serve as an upper-bound on the complexity of what is being asked of students to complete. This is not a theory oriented course, and is aimed at students looking to obtain practical knowledge and ultimately write code themselves in the future. The assignments assume that students do not have access to a GPU for days at a time, and so are designed that they can be completed within a day when done correctly. This constraint is born of insufficient funds to purchase GPUs for every student, while also not wanting to burden students with a large capital cost of a GPU when they do not yet know if they will enjoy deep learning. 

Each homework assignment consisted of 4 or 5 coding questions with concise summaries in Table \ref{tbl:qs}. Tasks included implementing feature processing, specific neural network architectures, and making specific modifications to an architecture in the book assigned, and comparing the impact of hyper-parameters on total run-time or accuracy of a model. 

\begin{table}[!h]
\caption{Concise descriptions of the kinds of tasks each homework question(s) required of the students. Each is built from one or more problems from the book written for the course used in this study, Inside Deep Learning. The chapter and question of the full content under the problems column.} \label{tbl:qs}
\begin{tabularx}{\columnwidth}{@{}clX@{}}
\toprule
HW & Problems & Task                                                                \\ \midrule
1  & C2, Q2   & Evaluate a model via AUC.                                            \\
1  & C2, Q3-4 & Implement checkpointed training.                                     \\
1  & C2, Q5   & Add more layers to a model                                          \\
2  & C3, Q2   & Train a CNN on CIFAR10                                              \\
2  & C4, Q1-3 & Train an RNN over text with a custom vocabulary.                     \\
3  & C5, Q6   & Perform hyperparameter tuning of a CNN.                             \\
3  & C6, Q2   & Train a deeper CNN with BatchNorm.                                  \\
3  & C6, Q6   & Train an LSTM and compare to an RNN.                                \\
3  & C7, Q1   & Train and auto-encoder on MNIST without classes 5 \& 9, then evaluate on the missing and included classes.              \\
3  & C7, Q8   & Create an autoregressive loader aware of sentence boundaries.      \\
4  & C8, Q4   & Replace pooling with strided convolutions in a U-Net                \\
4  & C9, Q1-2 & Implement a convolutional GAN.                                      \\
4  & C10, Q1  & Combine convolutions and attention for sequential image prediction. \\ \bottomrule
\end{tabularx}%
\end{table}

The first and fourth (last) homework were designed to be easier to complete. The first to avoid overwhelming students at the onset, and the second to allow students more time to work on a semester-long final project. There were generally two weeks between each homework assignment and due date, with the next homework being assigned the day the previous was due. The third homework was intentionally designed to be harder, and the instructors suspected students would not give themselves sufficient effort to complete the assignment. For this reason, a 1-2 week extension was baked into the curricula and used every semester. For this reason, student start times can be significantly larger for the third homework. 

The course policy included a ``no questions asked'' late grading policy, that allowed students to submit an assignment up to 72 hours late, for -10 points for each day the assignment was late. This meant a total late penalty of -30 hours was possible. This penalty was excluded from the data and calculations, as our goal is to make inferences about the value of additional time spent. 

Because all courses are different, our results can not be used to infer that every deep learning assignment, class, or set of exercises will follow the results of this paper. Indeed, this will always be the case for any course taught, and no singular study can infer a recommendation appropriate to all universities and classes. Our hope is this study will be informative and help encourage others to study this aspect of education and determine if the results may be applicable and informative to their own instruction. 

\subsection{Removed Records}

Not all student records were kept/used for this study. In total 7 student records (leaving 69 remaining) are excluded from our analysis due to the following:

\begin{itemize}
    \item The student did not follow requirements on making the submission editable or starting the homework in Colab, meaning we did not have access to the needed information. 
    \item The student cheated on the homework assignments, making them non-reflective of start time on student grades\footnote{Start time potentially impacted propensity to engage in academic misconduct, amongst other stressors with the pandemic. These considerations are critical but beyond our scope and data.}. 
    \item Catastrophic life event such as death of an immediate family member or significant change in medical status. 
\end{itemize}

The final project of the course is excluded from this study. Our experience was that students used multiple Colab instances in clever ways to make further progress on their final projects. This includes running multiple Colabs simultaneously to perform more experiments or using one to run experiments and a second to develop new code. Further, students were allowed to work with external companies/entities on their projects as a means of motivation and to provide more real-world experience (e.g., writing code in support of a favorite charity), and so was not always feasible to perform in a Colab environment. This made data collection on ``start'' times mostly meaningless, and further exacerbates the importance of true total time spent writing code over the gap between start and submission.

\subsection{Institutional Review Board Approval}

Our study considers human subjects (our students), and so was required to go through an Institutional Review Board (IRB) for approval. Our study was approved by the IRB based on two key factors: 1) Our study's design did not result in any change to student's grades, and was purely observational. 2) Our study did not infringe on any student's rights to privacy. 

The later point is particularly important in consideration of limitations in the study's results. As we have stated and will emphasize again, our data does not reflect a granular measure of effort or time spent. We are thus unable to differentiate between a student starting early and spending only a few minutes a day, versus a second student who started later but spent the same amount of time in a single session, to complete the assignment. Getting information at a more granular level would not be passed by our IRB in discussion with them, and we will exemplify two common questions we have received that are not satisfiable by our IRB. 

First, one may suggest that the students must engage in the use of some kind of version control system, with positive or negative incentives for frequent use of pushing code changes, such that the total time spent could be inferred from the edit history. This would be a noisy inference, but also would cause the study design to affect student grades or their perception of how they are graded. For this reason our IRB would not approve of a study with this kind of design. 

Second, it has been suggested that students should be monitored continuously to track when and for exactly how long they are performing the assignments. Beyond the logistical difficulties of monitoring $\approx$25 students over several months for multiple semesters, this imposes a significant invasion of the student's privacy. Students have an expectation of privacy outside of the classroom, and the monitoring required not only violates that privacy, but is plausibly illegal. This avenue is thus also ill-advised. 

For these reasons we find our approach satisficing in allowing us to study the question of interest, at a known and acknowledged level of imprecision. It is logistically feasible, does not alter student grades, and does not infringe on student's privacy rights.

\section{Model} \label{sec:model}
\begin{figure*}
    \centering
    \begin{adjustbox}{max width=0.9\textwidth}
    \includegraphics[]{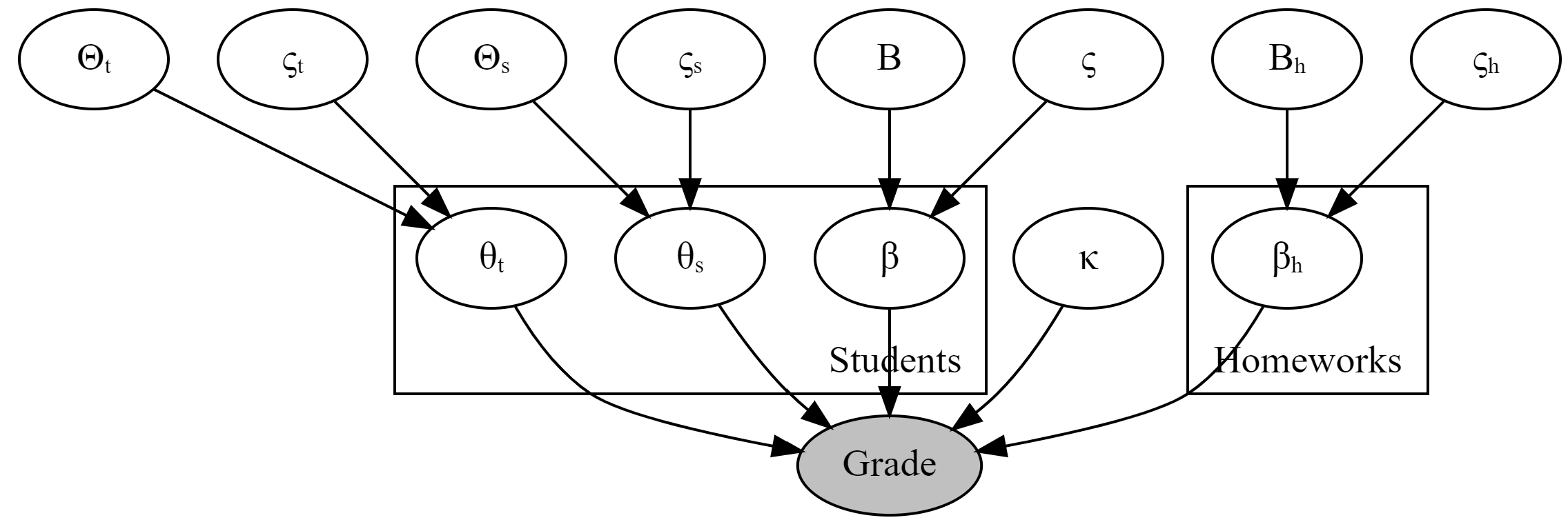}
   \end{adjustbox}
    \caption{Plate diagram of the variables used in our model. The box in a plate diagram indicates a repeated measure, as each student receives their own individual bias $\beta$ (how well does each individual student perform at baseline) and their own coefficients $\theta_t$ and $\theta_s$ describe their individual benefits from time spent on an assignment and submitting early. Similarly, each homework has a special bias term to quantify if certain homeworks required more time to complete than others. The top level hyper-priors describe the population level averages ($\Theta_t$, $\Theta_s$, $B$, $B_h$) and variances ($\varsigma_t$, $\varsigma_s$,  $\varsigma$, $\varsigma_h$) for the variables they point to. The variance for the beta regression $\kappa$ is inferred on its own as a property of the dataset as a whole. }
    \label{fig:plate_diagram}
\end{figure*}

To study the impact of student start and ``total'' time spent on assignments toward the grade received, we will use a linear hierarchical Bayesian model. The overall plate diagram is given in \cref{fig:plate_diagram}, and the generative story in \cref{alg:algorithm}. Capital Greek letters are used for hyper-priors and lower-case Greek letters for the priors. We use this hierarchical design because of the limited total amount of data, and our assumption is that there is shared information between students in behavior---but some are unique and should be modeled in such a way. Using a hierarchical model allows information sharing to occur across students and homeworks, while simultaneously allowing for variation between them ~\cite{Gelman2013}.  At a high level, our model incorperates the following design factors with further explenation after. 

\begin{enumerate}
    \item A hierarchical linear model is used to follow best practices to incorporate information sharing (e.g., students working on the same assignment). 
    \item Each student has an independent bias term, which allows the model to account for intrinsic differences in capability to complete the assignment (regardless of the source of those differences, e.g., innate ability or prior exposure).
    \item Using the population level hyper-prior to infer population level rates, and the per-student prior to allow for handling of per-student variance. 
\end{enumerate}

We will now detail the variables in our model and the logic behind their design. It also allows us to estimate credible intervals, that are a quantified estimate about the uncertainty of each hyper-prior (i.e., population level) and sample prior (i.e., student/homework level) to determine if there is a significant relationship, without being over-encumbered by multiple-test corrections increasing Type II errors in a more frequentist approach \cite{Gelman2012}
. We will use the heavy tailed Cauchy distribution for all hyper-priors as it imposes minimal assumption of the population level values, and a Gaussian distribution for other priors as a  reasonable default choice and we do not desire a heavy tail for the coefficients sampled from them.

\begin{algorithm}[!h]
\caption{Generative Story}
\label{alg:algorithm}
\textbf{Input}: Student start and total time spent on an assignment $x_t$ and $x_s$ for all students (index by $j$ superscript).
\begin{algorithmic}[1] %
\STATE $\Theta_t, \Theta_s, B, B_h \sim \mathrm{Cauchy}(0, 1)$ \COMMENT{Location hyper-priors} 
\STATE $\varsigma_t, \varsigma_s, \varsigma, \varsigma_h, \kappa \sim \mathrm{Cauchy}(0, 1)^+$ \COMMENT{Variance hyper-priors, truncated to non-negative values} 
\FOR{Each Homework $i$} 
    \STATE {$\beta_{h}^{i} \sim \mathcal{N}(B_h,\varsigma_h)$} \COMMENT{Each assignment gets a bias adjustment for difficulty}
\ENDFOR
\FOR{Each Student $j$} 
    \STATE {$\beta^{j} \sim \mathcal{N}(B,\varsigma)$} 
    \STATE {$\theta_{t}^{j} \sim \mathcal{N}(\Theta_t,\varsigma_t)$}
    \STATE {$\theta_{s}^{j} \sim \mathcal{N}(\Theta_s,\varsigma_s)$}
\ENDFOR
\STATE $\hat{\mu}^{i,j} \gets \sigma \left( \theta_{t}^{j} \cdot x_t^j + \theta_{s}^{j} \cdot x_s^j + \beta^j + \beta_h^i \right)$
\STATE{measure likelihood against observed grades with \cref{eq:beta_prop} using $\mu \gets \hat{\mu}$ and concentration $\kappa \gets \kappa$}
\end{algorithmic}
\end{algorithm}

First, we observe that some students often require less time than others to complete an assignment, and so we feel it would be inappropriate to use a single bias term. For this reason our model allows each of the $j$ students to have their own bias term $\beta^j$, determined by its hyper-prior $B$. Similarly, the homeworks were designed with the intention that the first and last should be easier than the others. So we include additional homework-specific bias terms $\beta^i_h$ for each of the $i$ different homework assignments.

Using $x^j_t$ and $x^j_s$ to denote the $j$th's student total time and starting time respectively, we will have corresponding covariance $\theta^j_s$ and $\theta^j_t$. Again these are student-specific, so that we may study if individual students benefit differently from having more time to work on an assignment. The means of the hyper-priors for these two covariates are then $\Theta_s$ and $\Theta_t$, and we will look to the hyper-prior posterior after inference to answer the question: \textit{do students at large benefit from more time spent on assignments}. Looking at the student specific $\theta^j_s$ and $\theta^j_t$ then tells us if students vary in their benefit of more time. 

For all hyper-priors (upper-Greek) we sample the mean from a Cauchy distribution, and the variance parameter from the zero truncated Cauchy. This is done to impose little constraint on the location and variance of the hyper-prior. The prior variables (lower-Greek) are samples from Gaussian distributions $\mathcal{N}($mean, variance$)$ based on the hyper-priors. 

The response variable of our model is treated as a Beta regression, and we use the proportional beta formulation as defined by \cref{eq:beta_prop} that allows us to specify a mean $\mu$ and non-negative variance $\kappa$ as it is easier to model\footnote{in this context $B$ is the beta function, and is not used in this context anywhere else in the manuscript}.

\begin{equation} \label{eq:beta_prop}
    \mathrm{Beta\_Prop}(\theta|\mu,\kappa) =
\frac{\theta^{\mu\kappa - 1} \, (1 - \theta)^{(1 - \mu)\kappa- 1}}{\mathrm{B}(\mu \kappa, (1 - \mu) \kappa)}
\end{equation}

We use the Beta regression model as it is a popular means of regressing over $(0,1)$ constrained response variables and fits our grade distribution. We prefer to clip the maximum grade of $1.0$ (i.e., 100\%) to 99.9 and the minimum grade from 0.0 to 0.001, as adding a twice inflated Beta regression would result in a complex to specify model, and complicate analysis due to many 100\% grades in our dataset. Functionally a 99\% and 100\% grade are equal demonstrations of content mastery, but a zero-one inflated model would treat these as meaningfully different events. 
To constrain our regression $\hat{\mu}$ of \cref{eq:beta_prop} to the range $(0, 1)$, we use the common sigmoid function as defined 
by $\sigma(x) = \frac{1}{1+\exp(-x)}$.

This fully specifies our model of student grades and the impact that time, measured by start time and total time spent, impacts students' grades. We use the Numpyro library~\cite{Phan2019} and the NUTS sampler~\cite{JMLR:v15:hoffman14a} for our model's inference with 300 burn-in cycles and 600 samples after. This results in a $\hat{r}=1.00$ for every parameter of the model, indicating full convergence. 

\section{Results} \label{sec:results}

\begin{figure}[!h]
    \centering
    \input{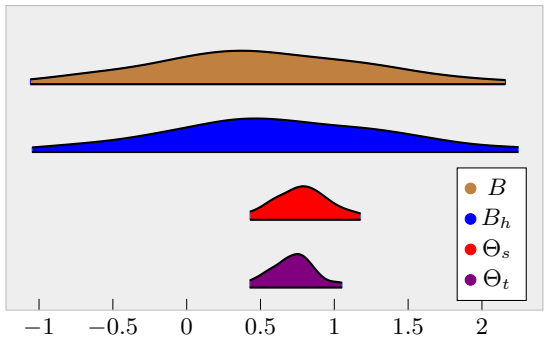}
    \caption{The 95\% credible interval of the hyper-priors $B$, the global bias term, and $B_h$, the homework-specific bias prior. $\Theta_s$ and $\Theta_t$ show the impact of submitting early and total time spent on an assignment; the overall result is a statistically significant positive correlation between these factors and receiving a higher grade.}
    \label{fig:hyper_prior}
\end{figure}

In this section we present our analysis of the results. We start with the posterior distribution of the hyper-priors shown with a 95\% credible interval, which can be found in \cref{fig:hyper_prior}. First are the global bias term $B$ and the homework specific bias prior $B_h$. In each case the wide distribution indicates the variability in student behavior. More notably $\Theta_s$ and $\Theta_t$ show the impact of submitting early and total time spent on an assignment respectively. In both cases the impact is positive and significant as zero is outside the credible interval, with mean values of 0.78 per week early submission and 0.72 per week of additional total time. We remind the reader that this week early corresponds to \textit{starting} the assignment a week early, and not a week of continuous effort.

Crucially, this allows us to examine questions about what the average student can do to improve their grade. One way to look at this is the rate of growth for the function:
$
\sigma(0.52+0.72\cdot x) - \sigma(0.52)
$.
This thus returns the impact of starting the assignment earlier, assuming the student will spend all available time to complete it (i.e, submits at the last minute the homework is due). In this form we can infer that starting $x=2$ weeks early instead of $x=1$ could yield a 10\% improvement in average grade received. This will invariably be affected by individual student performance, and so we must also ask about the distribution of individual students.

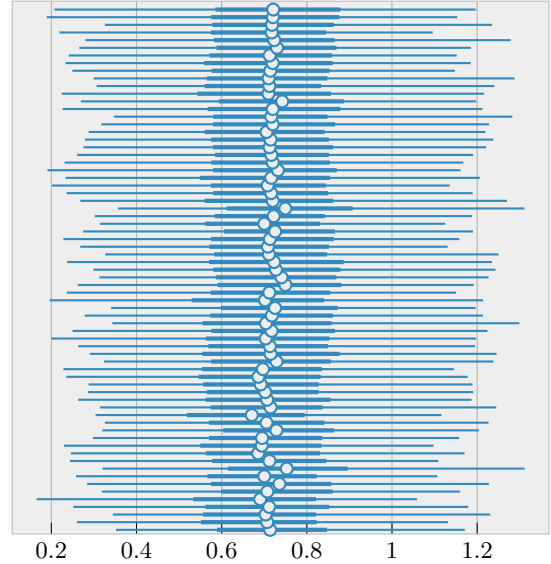
\begin{figure}[!h]
    \centering
    \input{figures/student-time-impact}
    \caption{Forest plot of the 95\% credible interval  of the student specific distributions $\theta_t$ that measure the impact of spending more total time on their final grade. Each line represents a different specific student, the the circle showing the median, thick blue lines showing the middle 50\% of the interval, and the thin blue lines showing the full 95\% credible interval. Results suggest a very consistent cross-population benefit to spending more total time on assignments.}
    \label{fig:student_total_time}
\end{figure}

Because $\Theta_s$ and $\Theta_t$ are hyper-priors over the student specific distributions of $\theta_s$ and $\theta_t$, we can look at these later distributions to understand the variability of impact. First we consider the total time spent on an assignment $\theta_t$ in \cref{fig:student_total_time}.

This plot shows the surprisingly consistent benefit that the student receives by spending a week's worth of time on each assignment. This would seem to imply that the benefits are stable and repeatable, and that given we model the problem with a sigmoid $\sigma$ we suspect corresponds to an implied diminishing return on the benefits of spending more time studying. This would also correspond to the data as shown in \cref{fig:data_view}, where after starting 200 hours (1.2 weeks) before the deadline all students obtain $\geq 85\%$ on their assignments.

\begin{figure}[!h]
    \centering
    \input{figures/student-start-impact}
    \caption{Forest plot of the 95\% credible interval  of the student specific distributions $\theta_s$ that measure the impact of submitting the homework earlier on individuals' grade. Each line represents a different specific student, with the circle showing the median, thick blue lines showing the middle 50\% of the interval, and the thin blue lines showing the full 95\% credible interval. Results suggest a consistent correlation between earlier submissions and improved outcomes.}
    \label{fig:student_submit_time}
\end{figure}
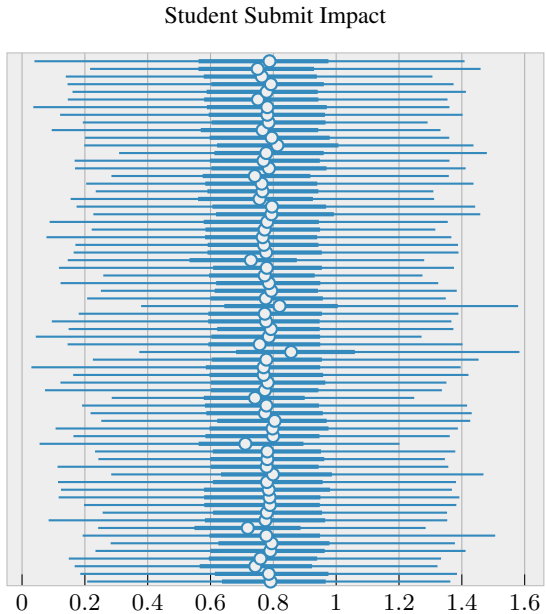

We note that the results in submission time are also highly consistent, as shown in \cref{fig:student_submit_time}. We suspect this is an effect that the submit time $x_s$ can be negative when students submitted late using the 72 hour late policy, as shown in \cref{fig:data_view}. Students \textit{starting} late never received a score better than a 70\% before the late penalty was applied. While not an original goal of our study, this does lead us to question the ultimate utility of the late submission policy. If removing the policy would encourage more students to start earlier, because the ``backup'' of using the late policy does not exist, we may obtain better total outcomes for all students. Simultaneously, our subjective feedback from student reviews and course evaluation is that the late policy is highly appreciated, and could lead to better performance via engagement. Answering this question is beyond the scope of this study, but an important point of future work identified by our data.

\begin{figure}[!h]
    \centering
    \input{figures/student-bias}
    \caption{Forest plot of the 95\% credible interval  of the student specific distributions $\beta$ that measure the student specific bias term on individual performance across all assignments. Each line represents a different specific student, the the circle showing the median, thick blue lines showing the middle 50\% of the interval, and the thin blue lines showing the full 95\% credible interval. Students' individual performance bias varies, suggesting that students vary significantly in effort required to complete assignments to a similar level.}
    \label{fig:bias}
\end{figure}
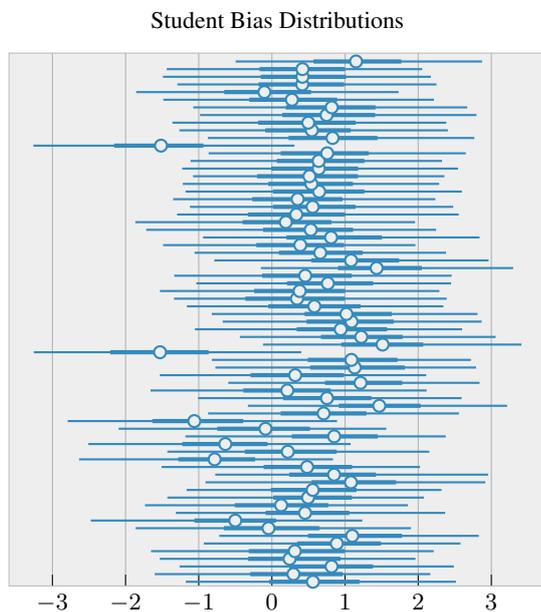

While submission and total time have stable distributions across students, students' individual performance biases display more asperity as shown in \cref{fig:bias}. In the more extreme cases two students had $\approx -2.5$ bias terms, placing them at a deficit of 3 weeks time compared to the mean student. In such a case this would require the student to start each homework assignment immediately in order to obtain the same outcomes. While technically feasible for the course as administered, this requires no other heavy work loads from the student's other courses throughout the semester, and is not realistically possible on all assignments. 

This result leads us to question what interventions may be possible to help such students. In our small study, these students are post-hoc identifiable by their performance on the first homework, where most students receive a perfect grade. We lack sufficient examples of such students to statistically confirm that this is reliably the case, but implies the possibility for early intervention may be possible. 

\section{Other Related Work} \label{sec:related_work}

In our introduction we reviewed two areas of education research that inspire and informed our study. We note that there is little other work regarding the broader question of time to complete or implement deep learning. There has been limited work in studying the amount of time students spend on coding assignments, but most studies are with respect to introductory computer science courses~\cite{segall2016much}. 
Empirical reproducibility work has used survival analysis to study the time it takes to replicate machine (and deep) learning papers, finding that most can be done quickly, but a long and heavy tail exists in the effort required~\cite{Raff2020c, Raff2019_quantify_repro,10.1145/3589806.3600042}. However, such work is focused on replicated academic peer-reviewed papers, rather than curated assignments in a course. That said, the gap between deep learning graduate course work and implementing academic papers is not too large, and while the annual ML Reproducibility Challenge \cite{Sinha2022} has demonstrated that students can succeed at such tasks, the additional information of time required has not been recorded. Adding such information to future years could prove valuable to reproducibility work and potentially inform the design of final course projects where attempting to replicate a paper is of a more appropriate scope. 

Towards potential interventions, previous work has found that explicitly teaching students how to debug their own code lessens the time they spend on coding assignments~\cite{10.1145/971300.971310}. It may be possible to develop similar interventions for deep learning, but most work in ``debugging'' deep learning is still oriented toward researchers\footnote{\url{https://debug-ml-iclr2019.github.io/}}, and it is not clear to us if the requisite stability in tools and techniques have been distilled to a level for students. We also note that while we are the first (to the best of our knowledge) to estimate time spent on an assignment from version history, we are not the first to use version history in relation to student assessment. In particular, \cite{10.1145/1082983.1083150} attempted to use version history to try and predict student outcomes on a given assignment, but found little predictive value. While such results could be improved today with better methods for predictive modeling, our more immediate concern would be on how to more precisely quantify actual time spent. Concerns about the use of Large Langage Models (LLMs) may also impact future results~\cite{10.1145/3511808.3557079,prather2023s,denny2023computing}, which was not a popular tool at the time this study was done. 

\section{Conclusion} \label{sec:conclusion}

We have conducted the first study of the impact of start time (i.e., starting a homework early) and the total time spent on a deep learning assignment on the grade received on the assignment. We find that both factors are statistically significantly correlated with improved scores by a student, which---while intuitive---is not identical to results found in other populations using different methodologies. The current data suggests that students do vary significantly in the base amount of time needed to complete an assignment, but the unit benefit of improvement is remarkably stable across students. This suggests that improved identification and intervention for students who require more time to complete assignments is worth further investigation. Our results suggest the average student may be able to improve their grade by 10\% by starting a week earlier than they would otherwise. 

Notably, our study is limited to students at a single institution, during three semesters of the COVID-19 pandemic. As such, our results may have selection bias and additional pandemic-related confounding factors. Simultaneously, the current environment suggests that we may unfortunately be operating in a future in which COVID is endemic, and so some of these biases may be relevant to future education concerns. Comparing results across curricula for courses at different institutions, and developing more precise methods of quantifying the actual total time spent, rather than our proxy measure from edit history, are directions for future work to improve upon. 

\bibliographystyle{ecai}
\bibliography{references,extra}

\end{document}

%% file: figures/student-time-impact.tex
\begin{tikzpicture}

\definecolor{darkgray178}{RGB}{178,178,178}
\definecolor{silver188}{RGB}{188,188,188}
\definecolor{steelblue52138189}{RGB}{52,138,189}
\definecolor{whitesmoke238}{RGB}{238,238,238}

\begin{axis}[
axis background/.style={fill=whitesmoke238},
axis line style={silver188},
title={Student Total Time Impact},
x grid style={darkgray178},
width=\columnwidth,
height=\columnwidth,
xmajorgrids,
xmin=0.107999596744776, xmax=1.3688581995666,
xtick pos=left,
xtick style={color=black},
y grid style={darkgray178},
ymajorticks=false,
ymin=-1.35, ymax=171,
ytick style={color=black},
ytick={0,2.475,4.95,7.425,9.9,12.375,14.85,17.325,19.8,22.275,24.75,27.225,29.7,32.175,34.65,37.125,39.6,42.075,44.55,47.025,49.5,51.975,54.45,56.925,59.4,61.875,64.35,66.825,69.3,71.775,74.25,76.725,79.2,81.675,84.15,86.625,89.1,91.575,94.05,96.5249999999999,98.9999999999999,101.475,103.95,106.425,108.9,111.375,113.85,116.325,118.8,121.275,123.75,126.225,128.7,131.175,133.65,136.125,138.6,141.075,143.55,146.025,148.5,150.975,153.45,155.925,158.4,160.875,163.35,165.825,168.3},
yticklabels={
  [68],
  [67],
  [66],
  [65],
  [64],
  [63],
  [62],
  [61],
  [60],
  [59],
  [58],
  [57],
  [56],
  [55],
  [54],
  [53],
  [52],
  [51],
  [50],
  [49],
  [48],
  [47],
  [46],
  [45],
  [44],
  [43],
  [42],
  [41],
  [40],
  [39],
  [38],
  [37],
  [36],
  [35],
  [34],
  [33],
  [32],
  [31],
  [30],
  [29],
  [28],
  [27],
  [26],
  [25],
  [24],
  [23],
  [22],
  [21],
  [20],
  [19],
  [18],
  [17],
  [16],
  [15],
  [14],
  [13],
  [12],
  [11],
  [10],
  [9],
  [8],
  [7],
  [6],
  [5],
  [4],
  [3],
  [2],
  [1],
  θₜ[0]
}
]
\path [draw=steelblue52138189, thick]
(axis cs:0.351963192224503,0)
--(axis cs:1.17095398902893,0);

\path [draw=steelblue52138189, ultra thick]
(axis cs:0.589131876826286,0)
--(axis cs:0.847490474581718,0);

\path [draw=steelblue52138189, thick]
(axis cs:0.259905397891998,2.475)
--(axis cs:1.13160681724548,2.475);

\path [draw=steelblue52138189, ultra thick]
(axis cs:0.551331982016563,2.475)
--(axis cs:0.822675243020058,2.475);

\path [draw=steelblue52138189, thick]
(axis cs:0.344328433275223,4.95)
--(axis cs:1.23082649707794,4.95);

\path [draw=steelblue52138189, ultra thick]
(axis cs:0.555988281965256,4.95)
--(axis cs:0.821398749947548,4.95);

\path [draw=steelblue52138189, thick]
(axis cs:0.251556515693665,7.425)
--(axis cs:1.17892503738403,7.425);

\path [draw=steelblue52138189, ultra thick]
(axis cs:0.562225058674812,7.425)
--(axis cs:0.852422162890434,7.425);

\path [draw=steelblue52138189, thick]
(axis cs:0.165311351418495,9.9)
--(axis cs:1.05867290496826,9.9);

\path [draw=steelblue52138189, ultra thick]
(axis cs:0.533642679452896,9.9)
--(axis cs:0.821439862251282,9.9);

\path [draw=steelblue52138189, thick]
(axis cs:0.318562090396881,12.375)
--(axis cs:1.15953636169434,12.375);

\path [draw=steelblue52138189, ultra thick]
(axis cs:0.599007427692413,12.375)
--(axis cs:0.860546290874481,12.375);

\path [draw=steelblue52138189, thick]
(axis cs:0.284121751785278,14.85)
--(axis cs:1.22733795642853,14.85);

\path [draw=steelblue52138189, ultra thick]
(axis cs:0.575030103325844,14.85)
--(axis cs:0.857507079839706,14.85);

\path [draw=steelblue52138189, thick]
(axis cs:0.257860273122787,17.325)
--(axis cs:1.10623669624329,17.325);

\path [draw=steelblue52138189, ultra thick]
(axis cs:0.566724553704262,17.325)
--(axis cs:0.822914332151413,17.325);

\path [draw=steelblue52138189, thick]
(axis cs:0.320091009140015,19.8)
--(axis cs:1.31154644489288,19.8);

\path [draw=steelblue52138189, ultra thick]
(axis cs:0.614642217755318,19.8)
--(axis cs:0.895949721336365,19.8);

\path [draw=steelblue52138189, thick]
(axis cs:0.243901655077934,22.275)
--(axis cs:1.10863280296326,22.275);

\path [draw=steelblue52138189, ultra thick]
(axis cs:0.577694028615952,22.275)
--(axis cs:0.845488056540489,22.275);

\path [draw=steelblue52138189, thick]
(axis cs:0.246013671159744,24.75)
--(axis cs:1.17044997215271,24.75);

\path [draw=steelblue52138189, ultra thick]
(axis cs:0.562642261385918,24.75)
--(axis cs:0.830884724855423,24.75);

\path [draw=steelblue52138189, thick]
(axis cs:0.229861706495285,27.225)
--(axis cs:1.0975649356842,27.225);

\path [draw=steelblue52138189, ultra thick]
(axis cs:0.549624219536781,27.225)
--(axis cs:0.83414064347744,27.225);

\path [draw=steelblue52138189, thick]
(axis cs:0.297651439905167,29.7)
--(axis cs:1.15780258178711,29.7);

\path [draw=steelblue52138189, ultra thick]
(axis cs:0.570222422480583,29.7)
--(axis cs:0.83572743833065,29.7);

\path [draw=steelblue52138189, thick]
(axis cs:0.319823533296585,32.175)
--(axis cs:1.20403504371643,32.175);

\path [draw=steelblue52138189, ultra thick]
(axis cs:0.603538766503334,32.175)
--(axis cs:0.864022016525269,32.175);

\path [draw=steelblue52138189, thick]
(axis cs:0.325695067644119,34.65)
--(axis cs:1.22669339179993,34.65);

\path [draw=steelblue52138189, ultra thick]
(axis cs:0.57051083445549,34.65)
--(axis cs:0.841263562440872,34.65);

\path [draw=steelblue52138189, thick]
(axis cs:0.303867876529694,37.125)
--(axis cs:1.11625981330872,37.125);

\path [draw=steelblue52138189, ultra thick]
(axis cs:0.518379226326942,37.125)
--(axis cs:0.794823631644249,37.125);

\path [draw=steelblue52138189, thick]
(axis cs:0.314019978046417,39.6)
--(axis cs:1.24478578567505,39.6);

\path [draw=steelblue52138189, ultra thick]
(axis cs:0.574205353856087,39.6)
--(axis cs:0.83664470911026,39.6);

\path [draw=steelblue52138189, thick]
(axis cs:0.262898236513138,42.075)
--(axis cs:1.18686151504517,42.075);

\path [draw=steelblue52138189, ultra thick]
(axis cs:0.562633782625198,42.075)
--(axis cs:0.85577167570591,42.075);

\path [draw=steelblue52138189, thick]
(axis cs:0.285812556743622,44.55)
--(axis cs:1.19099497795105,44.55);

\path [draw=steelblue52138189, ultra thick]
(axis cs:0.565213099122047,44.55)
--(axis cs:0.827014565467834,44.55);

\path [draw=steelblue52138189, thick]
(axis cs:0.287554383277893,47.025)
--(axis cs:1.18913590908051,47.025);

\path [draw=steelblue52138189, ultra thick]
(axis cs:0.556491643190384,47.025)
--(axis cs:0.827963560819626,47.025);

\path [draw=steelblue52138189, thick]
(axis cs:0.235108718276024,49.5)
--(axis cs:1.17768108844757,49.5);

\path [draw=steelblue52138189, ultra thick]
(axis cs:0.545826852321625,49.5)
--(axis cs:0.8324114382267,49.5);

\path [draw=steelblue52138189, thick]
(axis cs:0.228441178798676,51.975)
--(axis cs:1.14552104473114,51.975);

\path [draw=steelblue52138189, ultra thick]
(axis cs:0.553731545805931,51.975)
--(axis cs:0.835573092103004,51.975);

\path [draw=steelblue52138189, thick]
(axis cs:0.32387775182724,54.45)
--(axis cs:1.23808598518372,54.45);

\path [draw=steelblue52138189, ultra thick]
(axis cs:0.57563990354538,54.45)
--(axis cs:0.856368735432625,54.45);

\path [draw=steelblue52138189, thick]
(axis cs:0.290630459785461,56.925)
--(axis cs:1.24566423892975,56.925);

\path [draw=steelblue52138189, ultra thick]
(axis cs:0.554420083761215,56.925)
--(axis cs:0.877195104956627,56.925);

\path [draw=steelblue52138189, thick]
(axis cs:0.26327320933342,59.4)
--(axis cs:1.19531309604645,59.4);

\path [draw=steelblue52138189, ultra thick]
(axis cs:0.568712219595909,59.4)
--(axis cs:0.849685057997704,59.4);

\path [draw=steelblue52138189, thick]
(axis cs:0.199846997857094,61.875)
--(axis cs:1.19748544692993,61.875);

\path [draw=steelblue52138189, ultra thick]
(axis cs:0.562778636813164,61.875)
--(axis cs:0.852774947881699,61.875);

\path [draw=steelblue52138189, thick]
(axis cs:0.249748438596725,64.35)
--(axis cs:1.22413218021393,64.35);

\path [draw=steelblue52138189, ultra thick]
(axis cs:0.576036274433136,64.35)
--(axis cs:0.866178050637245,64.35);

\path [draw=steelblue52138189, thick]
(axis cs:0.343389898538589,66.825)
--(axis cs:1.29902529716492,66.825);

\path [draw=steelblue52138189, ultra thick]
(axis cs:0.554943904280663,66.825)
--(axis cs:0.858009681105614,66.825);

\path [draw=steelblue52138189, thick]
(axis cs:0.278490126132965,69.3)
--(axis cs:1.21345698833466,69.3);

\path [draw=steelblue52138189, ultra thick]
(axis cs:0.573324471712112,69.3)
--(axis cs:0.861407995223999,69.3);

\path [draw=steelblue52138189, thick]
(axis cs:0.340344160795212,71.775)
--(axis cs:1.19644200801849,71.775);

\path [draw=steelblue52138189, ultra thick]
(axis cs:0.598463371396065,71.775)
--(axis cs:0.872499391436577,71.775);

\path [draw=steelblue52138189, thick]
(axis cs:0.195951327681541,74.25)
--(axis cs:1.21358835697174,74.25);

\path [draw=steelblue52138189, ultra thick]
(axis cs:0.529858618974686,74.25)
--(axis cs:0.840546622872353,74.25);

\path [draw=steelblue52138189, thick]
(axis cs:0.236214682459831,76.725)
--(axis cs:1.15027356147766,76.725);

\path [draw=steelblue52138189, ultra thick]
(axis cs:0.574340209364891,76.725)
--(axis cs:0.85452564060688,76.725);

\path [draw=steelblue52138189, thick]
(axis cs:0.262055605649948,79.2)
--(axis cs:1.19179701805115,79.2);

\path [draw=steelblue52138189, ultra thick]
(axis cs:0.590425372123718,79.2)
--(axis cs:0.881156668066978,79.2);

\path [draw=steelblue52138189, thick]
(axis cs:0.312787652015686,81.675)
--(axis cs:1.22711527347565,81.675);

\path [draw=steelblue52138189, ultra thick]
(axis cs:0.586316600441933,81.675)
--(axis cs:0.869498118758202,81.675);

\path [draw=steelblue52138189, thick]
(axis cs:0.298702836036682,84.15)
--(axis cs:1.24309456348419,84.15);

\path [draw=steelblue52138189, ultra thick]
(axis cs:0.580346509814262,84.15)
--(axis cs:0.878808990120888,84.15);

\path [draw=steelblue52138189, thick]
(axis cs:0.237263083457947,86.625)
--(axis cs:1.23469805717468,86.625);

\path [draw=steelblue52138189, ultra thick]
(axis cs:0.570526495575905,86.625)
--(axis cs:0.886624857783318,86.625);

\path [draw=steelblue52138189, thick]
(axis cs:0.326758652925491,89.1)
--(axis cs:1.25026512145996,89.1);

\path [draw=steelblue52138189, ultra thick]
(axis cs:0.582521244883537,89.1)
--(axis cs:0.847156599164009,89.1);

\path [draw=steelblue52138189, thick]
(axis cs:0.26784360408783,91.575)
--(axis cs:1.13069212436676,91.575);

\path [draw=steelblue52138189, ultra thick]
(axis cs:0.570813715457916,91.575)
--(axis cs:0.85217273235321,91.575);

\path [draw=steelblue52138189, thick]
(axis cs:0.228333011269569,94.05)
--(axis cs:1.15815353393555,94.05);

\path [draw=steelblue52138189, ultra thick]
(axis cs:0.574870005249977,94.05)
--(axis cs:0.863226979970932,94.05);

\path [draw=steelblue52138189, thick]
(axis cs:0.274901568889618,96.5249999999999)
--(axis cs:1.19033265113831,96.5249999999999);

\path [draw=steelblue52138189, ultra thick]
(axis cs:0.605188399553299,96.5249999999999)
--(axis cs:0.865995511412621,96.5249999999999);

\path [draw=steelblue52138189, thick]
(axis cs:0.314462929964066,98.9999999999999)
--(axis cs:1.12476718425751,98.9999999999999);

\path [draw=steelblue52138189, ultra thick]
(axis cs:0.561264112591743,98.9999999999999)
--(axis cs:0.830860465764999,98.9999999999999);

\path [draw=steelblue52138189, thick]
(axis cs:0.301780432462692,101.475)
--(axis cs:1.18809580802917,101.475);

\path [draw=steelblue52138189, ultra thick]
(axis cs:0.583253264427185,101.475)
--(axis cs:0.842328399419785,101.475);

\path [draw=steelblue52138189, thick]
(axis cs:0.356442362070084,103.95)
--(axis cs:1.31082284450531,103.95);

\path [draw=steelblue52138189, ultra thick]
(axis cs:0.611391574144363,103.95)
--(axis cs:0.907525226473808,103.95);

\path [draw=steelblue52138189, thick]
(axis cs:0.268128275871277,106.425)
--(axis cs:1.27001929283142,106.425);

\path [draw=steelblue52138189, ultra thick]
(axis cs:0.55986376106739,106.425)
--(axis cs:0.861855253577232,106.425);

\path [draw=steelblue52138189, thick]
(axis cs:0.23556162416935,108.9)
--(axis cs:1.18941378593445,108.9);

\path [draw=steelblue52138189, ultra thick]
(axis cs:0.580118417739868,108.9)
--(axis cs:0.85027951002121,108.9);

\path [draw=steelblue52138189, thick]
(axis cs:0.201755106449127,111.375)
--(axis cs:1.13547253608704,111.375);

\path [draw=steelblue52138189, ultra thick]
(axis cs:0.574978351593018,111.375)
--(axis cs:0.844500169157982,111.375);

\path [draw=steelblue52138189, thick]
(axis cs:0.233251586556435,113.85)
--(axis cs:1.20629620552063,113.85);

\path [draw=steelblue52138189, ultra thick]
(axis cs:0.548613503575325,113.85)
--(axis cs:0.854208409786224,113.85);

\path [draw=steelblue52138189, thick]
(axis cs:0.190904557704926,116.325)
--(axis cs:1.16082751750946,116.325);

\path [draw=steelblue52138189, ultra thick]
(axis cs:0.579921618103981,116.325)
--(axis cs:0.870319679379463,116.325);

\path [draw=steelblue52138189, thick]
(axis cs:0.231386244297028,118.8)
--(axis cs:1.16731834411621,118.8);

\path [draw=steelblue52138189, ultra thick]
(axis cs:0.575542315840721,118.8)
--(axis cs:0.853861972689629,118.8);

\path [draw=steelblue52138189, thick]
(axis cs:0.260432034730911,121.275)
--(axis cs:1.19060719013214,121.275);

\path [draw=steelblue52138189, ultra thick]
(axis cs:0.584559619426727,121.275)
--(axis cs:0.851614817976952,121.275);

\path [draw=steelblue52138189, thick]
(axis cs:0.275249302387238,123.75)
--(axis cs:1.22127819061279,123.75);

\path [draw=steelblue52138189, ultra thick]
(axis cs:0.579180508852005,123.75)
--(axis cs:0.861244037747383,123.75);

\path [draw=steelblue52138189, thick]
(axis cs:0.279038727283478,126.225)
--(axis cs:1.23761820793152,126.225);

\path [draw=steelblue52138189, ultra thick]
(axis cs:0.574937149882317,126.225)
--(axis cs:0.851928740739822,126.225);

\path [draw=steelblue52138189, thick]
(axis cs:0.287728548049927,128.7)
--(axis cs:1.21922183036804,128.7);

\path [draw=steelblue52138189, ultra thick]
(axis cs:0.55983954668045,128.7)
--(axis cs:0.841575533151627,128.7);

\path [draw=steelblue52138189, thick]
(axis cs:0.317097455263138,131.175)
--(axis cs:1.2280021905899,131.175);

\path [draw=steelblue52138189, ultra thick]
(axis cs:0.579836249351501,131.175)
--(axis cs:0.866082459688187,131.175);

\path [draw=steelblue52138189, thick]
(axis cs:0.34720766544342,133.65)
--(axis cs:1.28265905380249,133.65);

\path [draw=steelblue52138189, ultra thick]
(axis cs:0.58029393851757,133.65)
--(axis cs:0.848562017083168,133.65);

\path [draw=steelblue52138189, thick]
(axis cs:0.226339295506477,136.125)
--(axis cs:1.21223723888397,136.125);

\path [draw=steelblue52138189, ultra thick]
(axis cs:0.566859677433968,136.125)
--(axis cs:0.8786401450634,136.125);

\path [draw=steelblue52138189, thick]
(axis cs:0.26936799287796,138.6)
--(axis cs:1.1976820230484,138.6);

\path [draw=steelblue52138189, ultra thick]
(axis cs:0.592828646302223,138.6)
--(axis cs:0.88710843026638,138.6);

\path [draw=steelblue52138189, thick]
(axis cs:0.224249735474586,141.075)
--(axis cs:1.21625876426697,141.075);

\path [draw=steelblue52138189, ultra thick]
(axis cs:0.543227300047874,141.075)
--(axis cs:0.856647923588753,141.075);

\path [draw=steelblue52138189, thick]
(axis cs:0.306097537279129,143.55)
--(axis cs:1.23997712135315,143.55);

\path [draw=steelblue52138189, ultra thick]
(axis cs:0.56031858921051,143.55)
--(axis cs:0.833819508552551,143.55);

\path [draw=steelblue52138189, thick]
(axis cs:0.299251079559326,146.025)
--(axis cs:1.28791642189026,146.025);

\path [draw=steelblue52138189, ultra thick]
(axis cs:0.56533944606781,146.025)
--(axis cs:0.848117828369141,146.025);

\path [draw=steelblue52138189, thick]
(axis cs:0.249182537198067,148.5)
--(axis cs:1.14769542217255,148.5);

\path [draw=steelblue52138189, ultra thick]
(axis cs:0.576169222593307,148.5)
--(axis cs:0.85328009724617,148.5);

\path [draw=steelblue52138189, thick]
(axis cs:0.233438089489937,150.975)
--(axis cs:1.18483698368073,150.975);

\path [draw=steelblue52138189, ultra thick]
(axis cs:0.559220969676971,150.975)
--(axis cs:0.861086130142212,150.975);

\path [draw=steelblue52138189, thick]
(axis cs:0.241266220808029,153.45)
--(axis cs:1.15203738212585,153.45);

\path [draw=steelblue52138189, ultra thick]
(axis cs:0.571383729577065,153.45)
--(axis cs:0.859072431921959,153.45);

\path [draw=steelblue52138189, thick]
(axis cs:0.26567530632019,155.925)
--(axis cs:1.18518316745758,155.925);

\path [draw=steelblue52138189, ultra thick]
(axis cs:0.587835282087326,155.925)
--(axis cs:0.869626834988594,155.925);

\path [draw=steelblue52138189, thick]
(axis cs:0.279940605163574,158.4)
--(axis cs:1.27848052978516,158.4);

\path [draw=steelblue52138189, ultra thick]
(axis cs:0.581792518496513,158.4)
--(axis cs:0.865359574556351,158.4);

\path [draw=steelblue52138189, thick]
(axis cs:0.218857452273369,160.875)
--(axis cs:1.09558165073395,160.875);

\path [draw=steelblue52138189, ultra thick]
(axis cs:0.574716776609421,160.875)
--(axis cs:0.845171570777893,160.875);

\path [draw=steelblue52138189, thick]
(axis cs:0.325727641582489,163.35)
--(axis cs:1.23484838008881,163.35);

\path [draw=steelblue52138189, ultra thick]
(axis cs:0.577688783407211,163.35)
--(axis cs:0.863532766699791,163.35);

\path [draw=steelblue52138189, thick]
(axis cs:0.189296498894691,165.825)
--(axis cs:1.15322852134705,165.825);

\path [draw=steelblue52138189, ultra thick]
(axis cs:0.575032874941826,165.825)
--(axis cs:0.876440927386284,165.825);

\path [draw=steelblue52138189, thick]
(axis cs:0.20696260035038,168.3)
--(axis cs:1.19633793830872,168.3);

\path [draw=steelblue52138189, ultra thick]
(axis cs:0.585258662700653,168.3)
--(axis cs:0.879182681441307,168.3);

\addplot [thick, steelblue52138189, mark=*, mark size=2.25, mark options={solid,fill=whitesmoke238}, only marks]
table {%
0.713832974433899 0
};
\addplot [thick, steelblue52138189, mark=*, mark size=2.25, mark options={solid,fill=whitesmoke238}, only marks]
table {%
0.706890493631363 2.475
};
\addplot [thick, steelblue52138189, mark=*, mark size=2.25, mark options={solid,fill=whitesmoke238}, only marks]
table {%
0.703382641077042 4.95
};
\addplot [thick, steelblue52138189, mark=*, mark size=2.25, mark options={solid,fill=whitesmoke238}, only marks]
table {%
0.71155908703804 7.425
};
\addplot [thick, steelblue52138189, mark=*, mark size=2.25, mark options={solid,fill=whitesmoke238}, only marks]
table {%
0.690415740013123 9.9
};
\addplot [thick, steelblue52138189, mark=*, mark size=2.25, mark options={solid,fill=whitesmoke238}, only marks]
table {%
0.707315415143967 12.375
};
\addplot [thick, steelblue52138189, mark=*, mark size=2.25, mark options={solid,fill=whitesmoke238}, only marks]
table {%
0.73559582233429 14.85
};
\addplot [thick, steelblue52138189, mark=*, mark size=2.25, mark options={solid,fill=whitesmoke238}, only marks]
table {%
0.70013439655304 17.325
};
\addplot [thick, steelblue52138189, mark=*, mark size=2.25, mark options={solid,fill=whitesmoke238}, only marks]
table {%
0.752858996391296 19.8
};
\addplot [thick, steelblue52138189, mark=*, mark size=2.25, mark options={solid,fill=whitesmoke238}, only marks]
table {%
0.712211936712265 22.275
};
\addplot [thick, steelblue52138189, mark=*, mark size=2.25, mark options={solid,fill=whitesmoke238}, only marks]
table {%
0.685759842395782 24.75
};
\addplot [thick, steelblue52138189, mark=*, mark size=2.25, mark options={solid,fill=whitesmoke238}, only marks]
table {%
0.694144934415817 27.225
};
\addplot [thick, steelblue52138189, mark=*, mark size=2.25, mark options={solid,fill=whitesmoke238}, only marks]
table {%
0.695042371749878 29.7
};
\addplot [thick, steelblue52138189, mark=*, mark size=2.25, mark options={solid,fill=whitesmoke238}, only marks]
table {%
0.728608161211014 32.175
};
\addplot [thick, steelblue52138189, mark=*, mark size=2.25, mark options={solid,fill=whitesmoke238}, only marks]
table {%
0.705330699682236 34.65
};
\addplot [thick, steelblue52138189, mark=*, mark size=2.25, mark options={solid,fill=whitesmoke238}, only marks]
table {%
0.670831978321075 37.125
};
\addplot [thick, steelblue52138189, mark=*, mark size=2.25, mark options={solid,fill=whitesmoke238}, only marks]
table {%
0.71484512090683 39.6
};
\addplot [thick, steelblue52138189, mark=*, mark size=2.25, mark options={solid,fill=whitesmoke238}, only marks]
table {%
0.706749469041824 42.075
};
\addplot [thick, steelblue52138189, mark=*, mark size=2.25, mark options={solid,fill=whitesmoke238}, only marks]
table {%
0.702298671007156 44.55
};
\addplot [thick, steelblue52138189, mark=*, mark size=2.25, mark options={solid,fill=whitesmoke238}, only marks]
table {%
0.691715866327286 47.025
};
\addplot [thick, steelblue52138189, mark=*, mark size=2.25, mark options={solid,fill=whitesmoke238}, only marks]
table {%
0.686136662960052 49.5
};
\addplot [thick, steelblue52138189, mark=*, mark size=2.25, mark options={solid,fill=whitesmoke238}, only marks]
table {%
0.696951568126678 51.975
};
\addplot [thick, steelblue52138189, mark=*, mark size=2.25, mark options={solid,fill=whitesmoke238}, only marks]
table {%
0.728870660066605 54.45
};
\addplot [thick, steelblue52138189, mark=*, mark size=2.25, mark options={solid,fill=whitesmoke238}, only marks]
table {%
0.714891374111176 56.925
};
\addplot [thick, steelblue52138189, mark=*, mark size=2.25, mark options={solid,fill=whitesmoke238}, only marks]
table {%
0.713323205709457 59.4
};
\addplot [thick, steelblue52138189, mark=*, mark size=2.25, mark options={solid,fill=whitesmoke238}, only marks]
table {%
0.703427225351334 61.875
};
\addplot [thick, steelblue52138189, mark=*, mark size=2.25, mark options={solid,fill=whitesmoke238}, only marks]
table {%
0.717210948467255 64.35
};
\addplot [thick, steelblue52138189, mark=*, mark size=2.25, mark options={solid,fill=whitesmoke238}, only marks]
table {%
0.703978687524796 66.825
};
\addplot [thick, steelblue52138189, mark=*, mark size=2.25, mark options={solid,fill=whitesmoke238}, only marks]
table {%
0.717863410711288 69.3
};
\addplot [thick, steelblue52138189, mark=*, mark size=2.25, mark options={solid,fill=whitesmoke238}, only marks]
table {%
0.7253497838974 71.775
};
\addplot [thick, steelblue52138189, mark=*, mark size=2.25, mark options={solid,fill=whitesmoke238}, only marks]
table {%
0.701738148927689 74.25
};
\addplot [thick, steelblue52138189, mark=*, mark size=2.25, mark options={solid,fill=whitesmoke238}, only marks]
table {%
0.711890518665314 76.725
};
\addplot [thick, steelblue52138189, mark=*, mark size=2.25, mark options={solid,fill=whitesmoke238}, only marks]
table {%
0.749684751033783 79.2
};
\addplot [thick, steelblue52138189, mark=*, mark size=2.25, mark options={solid,fill=whitesmoke238}, only marks]
table {%
0.740460723638535 81.675
};
\addplot [thick, steelblue52138189, mark=*, mark size=2.25, mark options={solid,fill=whitesmoke238}, only marks]
table {%
0.72754842042923 84.15
};
\addplot [thick, steelblue52138189, mark=*, mark size=2.25, mark options={solid,fill=whitesmoke238}, only marks]
table {%
0.72326648235321 86.625
};
\addplot [thick, steelblue52138189, mark=*, mark size=2.25, mark options={solid,fill=whitesmoke238}, only marks]
table {%
0.711058974266052 89.1
};
\addplot [thick, steelblue52138189, mark=*, mark size=2.25, mark options={solid,fill=whitesmoke238}, only marks]
table {%
0.708632558584213 91.575
};
\addplot [thick, steelblue52138189, mark=*, mark size=2.25, mark options={solid,fill=whitesmoke238}, only marks]
table {%
0.713233768939972 94.05
};
\addplot [thick, steelblue52138189, mark=*, mark size=2.25, mark options={solid,fill=whitesmoke238}, only marks]
table {%
0.725502461194992 96.5249999999999
};
\addplot [thick, steelblue52138189, mark=*, mark size=2.25, mark options={solid,fill=whitesmoke238}, only marks]
table {%
0.700176775455475 98.9999999999999
};
\addplot [thick, steelblue52138189, mark=*, mark size=2.25, mark options={solid,fill=whitesmoke238}, only marks]
table {%
0.722198814153671 101.475
};
\addplot [thick, steelblue52138189, mark=*, mark size=2.25, mark options={solid,fill=whitesmoke238}, only marks]
table {%
0.749384045600891 103.95
};
\addplot [thick, steelblue52138189, mark=*, mark size=2.25, mark options={solid,fill=whitesmoke238}, only marks]
table {%
0.719991266727448 106.425
};
\addplot [thick, steelblue52138189, mark=*, mark size=2.25, mark options={solid,fill=whitesmoke238}, only marks]
table {%
0.716154903173447 108.9
};
\addplot [thick, steelblue52138189, mark=*, mark size=2.25, mark options={solid,fill=whitesmoke238}, only marks]
table {%
0.707964718341827 111.375
};
\addplot [thick, steelblue52138189, mark=*, mark size=2.25, mark options={solid,fill=whitesmoke238}, only marks]
table {%
0.716231465339661 113.85
};
\addplot [thick, steelblue52138189, mark=*, mark size=2.25, mark options={solid,fill=whitesmoke238}, only marks]
table {%
0.732109934091568 116.325
};
\addplot [thick, steelblue52138189, mark=*, mark size=2.25, mark options={solid,fill=whitesmoke238}, only marks]
table {%
0.719999074935913 118.8
};
\addplot [thick, steelblue52138189, mark=*, mark size=2.25, mark options={solid,fill=whitesmoke238}, only marks]
table {%
0.716597020626068 121.275
};
\addplot [thick, steelblue52138189, mark=*, mark size=2.25, mark options={solid,fill=whitesmoke238}, only marks]
table {%
0.712661057710648 123.75
};
\addplot [thick, steelblue52138189, mark=*, mark size=2.25, mark options={solid,fill=whitesmoke238}, only marks]
table {%
0.714918255805969 126.225
};
\addplot [thick, steelblue52138189, mark=*, mark size=2.25, mark options={solid,fill=whitesmoke238}, only marks]
table {%
0.705376327037811 128.7
};
\addplot [thick, steelblue52138189, mark=*, mark size=2.25, mark options={solid,fill=whitesmoke238}, only marks]
table {%
0.71993300318718 131.175
};
\addplot [thick, steelblue52138189, mark=*, mark size=2.25, mark options={solid,fill=whitesmoke238}, only marks]
table {%
0.716447532176971 133.65
};
\addplot [thick, steelblue52138189, mark=*, mark size=2.25, mark options={solid,fill=whitesmoke238}, only marks]
table {%
0.719944417476654 136.125
};
\addplot [thick, steelblue52138189, mark=*, mark size=2.25, mark options={solid,fill=whitesmoke238}, only marks]
table {%
0.742127001285553 138.6
};
\addplot [thick, steelblue52138189, mark=*, mark size=2.25, mark options={solid,fill=whitesmoke238}, only marks]
table {%
0.709883749485016 141.075
};
\addplot [thick, steelblue52138189, mark=*, mark size=2.25, mark options={solid,fill=whitesmoke238}, only marks]
table {%
0.712038457393646 143.55
};
\addplot [thick, steelblue52138189, mark=*, mark size=2.25, mark options={solid,fill=whitesmoke238}, only marks]
table {%
0.710262149572372 146.025
};
\addplot [thick, steelblue52138189, mark=*, mark size=2.25, mark options={solid,fill=whitesmoke238}, only marks]
table {%
0.714191406965256 148.5
};
\addplot [thick, steelblue52138189, mark=*, mark size=2.25, mark options={solid,fill=whitesmoke238}, only marks]
table {%
0.719866186380386 150.975
};
\addplot [thick, steelblue52138189, mark=*, mark size=2.25, mark options={solid,fill=whitesmoke238}, only marks]
table {%
0.712635934352875 153.45
};
\addplot [thick, steelblue52138189, mark=*, mark size=2.25, mark options={solid,fill=whitesmoke238}, only marks]
table {%
0.730007886886597 155.925
};
\addplot [thick, steelblue52138189, mark=*, mark size=2.25, mark options={solid,fill=whitesmoke238}, only marks]
table {%
0.723465532064438 158.4
};
\addplot [thick, steelblue52138189, mark=*, mark size=2.25, mark options={solid,fill=whitesmoke238}, only marks]
table {%
0.717601180076599 160.875
};
\addplot [thick, steelblue52138189, mark=*, mark size=2.25, mark options={solid,fill=whitesmoke238}, only marks]
table {%
0.71833524107933 163.35
};
\addplot [thick, steelblue52138189, mark=*, mark size=2.25, mark options={solid,fill=whitesmoke238}, only marks]
table {%
0.720475167036057 165.825
};
\addplot [thick, steelblue52138189, mark=*, mark size=2.25, mark options={solid,fill=whitesmoke238}, only marks]
table {%
0.721063017845154 168.3
};
\end{axis}

\end{tikzpicture}

%% file: figures/student-start-impact.tex
\begin{tikzpicture}

\definecolor{darkgray178}{RGB}{178,178,178}
\definecolor{silver188}{RGB}{188,188,188}
\definecolor{steelblue52138189}{RGB}{52,138,189}
\definecolor{whitesmoke238}{RGB}{238,238,238}

\begin{axis}[
axis background/.style={fill=whitesmoke238},
axis line style={silver188},
title={Student Submit Impact},
x grid style={darkgray178},
width=\columnwidth,
height=\columnwidth,
xmajorgrids,
xmin=-0.0479778228327632, xmax=1.66115523036569,
xtick pos=left,
xtick style={color=black},
y grid style={darkgray178},
ymajorticks=false,
ymin=-1.35, ymax=171,
ytick style={color=black},
ytick={0,2.475,4.95,7.425,9.9,12.375,14.85,17.325,19.8,22.275,24.75,27.225,29.7,32.175,34.65,37.125,39.6,42.075,44.55,47.025,49.5,51.975,54.45,56.925,59.4,61.875,64.35,66.825,69.3,71.775,74.25,76.725,79.2,81.675,84.15,86.625,89.1,91.575,94.05,96.5249999999999,98.9999999999999,101.475,103.95,106.425,108.9,111.375,113.85,116.325,118.8,121.275,123.75,126.225,128.7,131.175,133.65,136.125,138.6,141.075,143.55,146.025,148.5,150.975,153.45,155.925,158.4,160.875,163.35,165.825,168.3},
yticklabels={
  [68],
  [67],
  [66],
  [65],
  [64],
  [63],
  [62],
  [61],
  [60],
  [59],
  [58],
  [57],
  [56],
  [55],
  [54],
  [53],
  [52],
  [51],
  [50],
  [49],
  [48],
  [47],
  [46],
  [45],
  [44],
  [43],
  [42],
  [41],
  [40],
  [39],
  [38],
  [37],
  [36],
  [35],
  [34],
  [33],
  [32],
  [31],
  [30],
  [29],
  [28],
  [27],
  [26],
  [25],
  [24],
  [23],
  [22],
  [21],
  [20],
  [19],
  [18],
  [17],
  [16],
  [15],
  [14],
  [13],
  [12],
  [11],
  [10],
  [9],
  [8],
  [7],
  [6],
  [5],
  [4],
  [3],
  [2],
  [1],
  θₛ[0]
}
]
\path [draw=steelblue52138189, thick]
(axis cs:0.207384198904037,0)
--(axis cs:1.36346936225891,0);

\path [draw=steelblue52138189, ultra thick]
(axis cs:0.636436134576797,0)
--(axis cs:0.966682955622673,0);

\path [draw=steelblue52138189, thick]
(axis cs:0.185406029224396,2.475)
--(axis cs:1.38458168506622,2.475);

\path [draw=steelblue52138189, ultra thick]
(axis cs:0.613785371184349,2.475)
--(axis cs:0.975325211882591,2.475);

\path [draw=steelblue52138189, thick]
(axis cs:0.167569041252136,4.95)
--(axis cs:1.32262790203094,4.95);

\path [draw=steelblue52138189, ultra thick]
(axis cs:0.565583974123001,4.95)
--(axis cs:0.923168614506721,4.95);

\path [draw=steelblue52138189, thick]
(axis cs:0.148735716938972,7.425)
--(axis cs:1.33386993408203,7.425);

\path [draw=steelblue52138189, ultra thick]
(axis cs:0.594485864043236,7.425)
--(axis cs:0.939519047737122,7.425);

\path [draw=steelblue52138189, thick]
(axis cs:0.233777284622192,9.9)
--(axis cs:1.41187620162964,9.9);

\path [draw=steelblue52138189, ultra thick]
(axis cs:0.598812460899353,9.9)
--(axis cs:0.963529765605927,9.9);

\path [draw=steelblue52138189, thick]
(axis cs:0.282086491584778,12.375)
--(axis cs:1.37804698944092,12.375);

\path [draw=steelblue52138189, ultra thick]
(axis cs:0.624861910939217,12.375)
--(axis cs:0.978636294603348,12.375);

\path [draw=steelblue52138189, thick]
(axis cs:0.192926988005638,14.85)
--(axis cs:1.50625777244568,14.85);

\path [draw=steelblue52138189, ultra thick]
(axis cs:0.596475407481194,14.85)
--(axis cs:0.948397934436798,14.85);

\path [draw=steelblue52138189, thick]
(axis cs:0.242377936840057,17.325)
--(axis cs:1.28488409519196,17.325);

\path [draw=steelblue52138189, ultra thick]
(axis cs:0.548849359154701,17.325)
--(axis cs:0.886631160974503,17.325);

\path [draw=steelblue52138189, thick]
(axis cs:0.0846467316150665,19.8)
--(axis cs:1.35356533527374,19.8);

\path [draw=steelblue52138189, ultra thick]
(axis cs:0.582108393311501,19.8)
--(axis cs:0.965025871992111,19.8);

\path [draw=steelblue52138189, thick]
(axis cs:0.256555408239365,22.275)
--(axis cs:1.35292446613312,22.275);

\path [draw=steelblue52138189, ultra thick]
(axis cs:0.608097821474075,22.275)
--(axis cs:0.955108776688576,22.275);

\path [draw=steelblue52138189, thick]
(axis cs:0.197186037898064,24.75)
--(axis cs:1.38249123096466,24.75);

\path [draw=steelblue52138189, ultra thick]
(axis cs:0.580069363117218,24.75)
--(axis cs:0.947834551334381,24.75);

\path [draw=steelblue52138189, thick]
(axis cs:0.116581536829472,27.225)
--(axis cs:1.39220309257507,27.225);

\path [draw=steelblue52138189, ultra thick]
(axis cs:0.579626083374023,27.225)
--(axis cs:0.949735820293427,27.225);

\path [draw=steelblue52138189, thick]
(axis cs:0.124068699777126,29.7)
--(axis cs:1.36847412586212,29.7);

\path [draw=steelblue52138189, ultra thick]
(axis cs:0.57924585044384,29.7)
--(axis cs:0.980529621243477,29.7);

\path [draw=steelblue52138189, thick]
(axis cs:0.114451013505459,32.175)
--(axis cs:1.38140904903412,32.175);

\path [draw=steelblue52138189, ultra thick]
(axis cs:0.607471898198128,32.175)
--(axis cs:0.956757768988609,32.175);

\path [draw=steelblue52138189, thick]
(axis cs:0.283934026956558,34.65)
--(axis cs:1.46885573863983,34.65);

\path [draw=steelblue52138189, ultra thick]
(axis cs:0.633384093642235,34.65)
--(axis cs:0.986567050218582,34.65);

\path [draw=steelblue52138189, thick]
(axis cs:0.113033339381218,37.125)
--(axis cs:1.3570042848587,37.125);

\path [draw=steelblue52138189, ultra thick]
(axis cs:0.6004329174757,37.125)
--(axis cs:0.944092899560928,37.125);

\path [draw=steelblue52138189, thick]
(axis cs:0.242925316095352,39.6)
--(axis cs:1.3461879491806,39.6);

\path [draw=steelblue52138189, ultra thick]
(axis cs:0.598575413227081,39.6)
--(axis cs:0.962074384093285,39.6);

\path [draw=steelblue52138189, thick]
(axis cs:0.232344686985016,42.075)
--(axis cs:1.37916231155396,42.075);

\path [draw=steelblue52138189, ultra thick]
(axis cs:0.607266023755074,42.075)
--(axis cs:0.951660096645355,42.075);

\path [draw=steelblue52138189, thick]
(axis cs:0.0558097772300243,44.55)
--(axis cs:1.20150172710419,44.55);

\path [draw=steelblue52138189, ultra thick]
(axis cs:0.562080174684525,44.55)
--(axis cs:0.895859181880951,44.55);

\path [draw=steelblue52138189, thick]
(axis cs:0.163611188530922,47.025)
--(axis cs:1.36150145530701,47.025);

\path [draw=steelblue52138189, ultra thick]
(axis cs:0.5837522149086,47.025)
--(axis cs:0.946784719824791,47.025);

\path [draw=steelblue52138189, thick]
(axis cs:0.106763504445553,49.5)
--(axis cs:1.3874124288559,49.5);

\path [draw=steelblue52138189, ultra thick]
(axis cs:0.596640855073929,49.5)
--(axis cs:0.975547954440117,49.5);

\path [draw=steelblue52138189, thick]
(axis cs:0.252190560102463,51.975)
--(axis cs:1.42677056789398,51.975);

\path [draw=steelblue52138189, ultra thick]
(axis cs:0.621355623006821,51.975)
--(axis cs:0.969736948609352,51.975);

\path [draw=steelblue52138189, thick]
(axis cs:0.218134984374046,54.45)
--(axis cs:1.43152582645416,54.45);

\path [draw=steelblue52138189, ultra thick]
(axis cs:0.586601406335831,54.45)
--(axis cs:0.957337856292725,54.45);

\path [draw=steelblue52138189, thick]
(axis cs:0.190985530614853,56.925)
--(axis cs:1.41676986217499,56.925);

\path [draw=steelblue52138189, ultra thick]
(axis cs:0.582287281751633,56.925)
--(axis cs:0.945269405841827,56.925);

\path [draw=steelblue52138189, thick]
(axis cs:0.285512715578079,59.4)
--(axis cs:1.24856603145599,59.4);

\path [draw=steelblue52138189, ultra thick]
(axis cs:0.578613981604576,59.4)
--(axis cs:0.899814933538437,59.4);

\path [draw=steelblue52138189, thick]
(axis cs:0.0730231404304504,61.875)
--(axis cs:1.33664882183075,61.875);

\path [draw=steelblue52138189, ultra thick]
(axis cs:0.601514488458633,61.875)
--(axis cs:0.943399116396904,61.875);

\path [draw=steelblue52138189, thick]
(axis cs:0.12229111790657,64.35)
--(axis cs:1.35099852085114,64.35);

\path [draw=steelblue52138189, ultra thick]
(axis cs:0.601242035627365,64.35)
--(axis cs:0.965555354952812,64.35);

\path [draw=steelblue52138189, thick]
(axis cs:0.16286589205265,66.825)
--(axis cs:1.42113077640533,66.825);

\path [draw=steelblue52138189, ultra thick]
(axis cs:0.596515819430351,66.825)
--(axis cs:0.958456829190254,66.825);

\path [draw=steelblue52138189, thick]
(axis cs:0.0297100432217121,69.3)
--(axis cs:1.39639186859131,69.3);

\path [draw=steelblue52138189, ultra thick]
(axis cs:0.584986612200737,69.3)
--(axis cs:0.948896035552025,69.3);

\path [draw=steelblue52138189, thick]
(axis cs:0.225389033555984,71.775)
--(axis cs:1.4533953666687,71.775);

\path [draw=steelblue52138189, ultra thick]
(axis cs:0.603418961167336,71.775)
--(axis cs:0.95449835062027,71.775);

\path [draw=steelblue52138189, thick]
(axis cs:0.373296439647675,74.25)
--(axis cs:1.58346736431122,74.25);

\path [draw=steelblue52138189, ultra thick]
(axis cs:0.681332439184189,74.25)
--(axis cs:1.05947712063789,74.25);

\path [draw=steelblue52138189, thick]
(axis cs:0.145070374011993,76.725)
--(axis cs:1.40238177776337,76.725);

\path [draw=steelblue52138189, ultra thick]
(axis cs:0.59323163330555,76.725)
--(axis cs:0.949239701032639,76.725);

\path [draw=steelblue52138189, thick]
(axis cs:0.0437426753342152,79.2)
--(axis cs:1.27182078361511,79.2);

\path [draw=steelblue52138189, ultra thick]
(axis cs:0.602589547634125,79.2)
--(axis cs:0.948837444186211,79.2);

\path [draw=steelblue52138189, thick]
(axis cs:0.148198530077934,81.675)
--(axis cs:1.37303340435028,81.675);

\path [draw=steelblue52138189, ultra thick]
(axis cs:0.621035665273666,81.675)
--(axis cs:0.947662308812141,81.675);

\path [draw=steelblue52138189, thick]
(axis cs:0.0949007794260979,84.15)
--(axis cs:1.36682629585266,84.15);

\path [draw=steelblue52138189, ultra thick]
(axis cs:0.59263177216053,84.15)
--(axis cs:0.947398066520691,84.15);

\path [draw=steelblue52138189, thick]
(axis cs:0.180455729365349,86.625)
--(axis cs:1.38919532299042,86.625);

\path [draw=steelblue52138189, ultra thick]
(axis cs:0.593114048242569,86.625)
--(axis cs:0.954528629779816,86.625);

\path [draw=steelblue52138189, thick]
(axis cs:0.379174679517746,89.1)
--(axis cs:1.57935440540314,89.1);

\path [draw=steelblue52138189, ultra thick]
(axis cs:0.643805265426636,89.1)
--(axis cs:1.00514683127403,89.1);

\path [draw=steelblue52138189, thick]
(axis cs:0.207238733768463,91.575)
--(axis cs:1.3488974571228,91.575);

\path [draw=steelblue52138189, ultra thick]
(axis cs:0.599625557661057,91.575)
--(axis cs:0.957696110010147,91.575);

\path [draw=steelblue52138189, thick]
(axis cs:0.250933945178986,94.05)
--(axis cs:1.3837902545929,94.05);

\path [draw=steelblue52138189, ultra thick]
(axis cs:0.613299801945686,94.05)
--(axis cs:0.941990166902542,94.05);

\path [draw=steelblue52138189, thick]
(axis cs:0.122643798589706,96.5249999999999)
--(axis cs:1.32493162155151,96.5249999999999);

\path [draw=steelblue52138189, ultra thick]
(axis cs:0.617819473147392,96.5249999999999)
--(axis cs:0.949318319559097,96.5249999999999);

\path [draw=steelblue52138189, thick]
(axis cs:0.258614897727966,98.9999999999999)
--(axis cs:1.27469420433044,98.9999999999999);

\path [draw=steelblue52138189, ultra thick]
(axis cs:0.595436170697212,98.9999999999999)
--(axis cs:0.931553304195404,98.9999999999999);

\path [draw=steelblue52138189, thick]
(axis cs:0.117492459714413,101.475)
--(axis cs:1.37499916553497,101.475);

\path [draw=steelblue52138189, ultra thick]
(axis cs:0.608224168419838,101.475)
--(axis cs:0.954540565609932,101.475);

\path [draw=steelblue52138189, thick]
(axis cs:0.145408511161804,103.95)
--(axis cs:1.28022086620331,103.95);

\path [draw=steelblue52138189, ultra thick]
(axis cs:0.533163920044899,103.95)
--(axis cs:0.874669253826141,103.95);

\path [draw=steelblue52138189, thick]
(axis cs:0.164242386817932,106.425)
--(axis cs:1.38930225372314,106.425);

\path [draw=steelblue52138189, ultra thick]
(axis cs:0.59037446975708,106.425)
--(axis cs:0.954264312982559,106.425);

\path [draw=steelblue52138189, thick]
(axis cs:0.169846400618553,108.9)
--(axis cs:1.38787078857422,108.9);

\path [draw=steelblue52138189, ultra thick]
(axis cs:0.591173321008682,108.9)
--(axis cs:0.943465456366539,108.9);

\path [draw=steelblue52138189, thick]
(axis cs:0.0777246057987213,111.375)
--(axis cs:1.36654305458069,111.375);

\path [draw=steelblue52138189, ultra thick]
(axis cs:0.582533329725266,111.375)
--(axis cs:0.939395919442177,111.375);

\path [draw=steelblue52138189, thick]
(axis cs:0.221747517585754,113.85)
--(axis cs:1.31564605236053,113.85);

\path [draw=steelblue52138189, ultra thick]
(axis cs:0.584036767482758,113.85)
--(axis cs:0.948620706796646,113.85);

\path [draw=steelblue52138189, thick]
(axis cs:0.0879380330443382,116.325)
--(axis cs:1.3551698923111,116.325);

\path [draw=steelblue52138189, ultra thick]
(axis cs:0.578297436237335,116.325)
--(axis cs:0.944844871759415,116.325);

\path [draw=steelblue52138189, thick]
(axis cs:0.227016806602478,118.8)
--(axis cs:1.45872855186462,118.8);

\path [draw=steelblue52138189, ultra thick]
(axis cs:0.617336094379425,118.8)
--(axis cs:0.992593318223953,118.8);

\path [draw=steelblue52138189, thick]
(axis cs:0.173509776592255,121.275)
--(axis cs:1.44246888160706,121.275);

\path [draw=steelblue52138189, ultra thick]
(axis cs:0.605933755636215,121.275)
--(axis cs:0.96790386736393,121.275);

\path [draw=steelblue52138189, thick]
(axis cs:0.154784634709358,123.75)
--(axis cs:1.31181263923645,123.75);

\path [draw=steelblue52138189, ultra thick]
(axis cs:0.560274541378021,123.75)
--(axis cs:0.925955161452293,123.75);

\path [draw=steelblue52138189, thick]
(axis cs:0.234603852033615,126.225)
--(axis cs:1.30934190750122,126.225);

\path [draw=steelblue52138189, ultra thick]
(axis cs:0.590042680501938,126.225)
--(axis cs:0.943390443921089,126.225);

\path [draw=steelblue52138189, thick]
(axis cs:0.204157680273056,128.7)
--(axis cs:1.43716239929199,128.7);

\path [draw=steelblue52138189, ultra thick]
(axis cs:0.583216726779938,128.7)
--(axis cs:0.939712136983871,128.7);

\path [draw=steelblue52138189, thick]
(axis cs:0.284554153680801,131.175)
--(axis cs:1.35866451263428,131.175);

\path [draw=steelblue52138189, ultra thick]
(axis cs:0.57489849627018,131.175)
--(axis cs:0.917853310704231,131.175);

\path [draw=steelblue52138189, thick]
(axis cs:0.168456107378006,133.65)
--(axis cs:1.41214692592621,133.65);

\path [draw=steelblue52138189, ultra thick]
(axis cs:0.602567598223686,133.65)
--(axis cs:0.96935972571373,133.65);

\path [draw=steelblue52138189, thick]
(axis cs:0.167279630899429,136.125)
--(axis cs:1.36071813106537,136.125);

\path [draw=steelblue52138189, ultra thick]
(axis cs:0.59808437526226,136.125)
--(axis cs:0.947985097765923,136.125);

\path [draw=steelblue52138189, thick]
(axis cs:0.308956652879715,138.6)
--(axis cs:1.47963035106659,138.6);

\path [draw=steelblue52138189, ultra thick]
(axis cs:0.612313568592072,138.6)
--(axis cs:0.960298240184784,138.6);

\path [draw=steelblue52138189, thick]
(axis cs:0.197951540350914,141.075)
--(axis cs:1.43702447414398,141.075);

\path [draw=steelblue52138189, ultra thick]
(axis cs:0.620646864175797,141.075)
--(axis cs:1.00703647732735,141.075);

\path [draw=steelblue52138189, thick]
(axis cs:0.200325980782509,143.55)
--(axis cs:1.35995006561279,143.55);

\path [draw=steelblue52138189, ultra thick]
(axis cs:0.597152456641197,143.55)
--(axis cs:0.979383215308189,143.55);

\path [draw=steelblue52138189, thick]
(axis cs:0.0944194942712784,146.025)
--(axis cs:1.33198642730713,146.025);

\path [draw=steelblue52138189, ultra thick]
(axis cs:0.568826362490654,146.025)
--(axis cs:0.942797690629959,146.025);

\path [draw=steelblue52138189, thick]
(axis cs:0.193356707692146,148.5)
--(axis cs:1.29096066951752,148.5);

\path [draw=steelblue52138189, ultra thick]
(axis cs:0.603287160396576,148.5)
--(axis cs:0.966041505336761,148.5);

\path [draw=steelblue52138189, thick]
(axis cs:0.121145099401474,150.975)
--(axis cs:1.40253281593323,150.975);

\path [draw=steelblue52138189, ultra thick]
(axis cs:0.594016015529633,150.975)
--(axis cs:0.964338093996048,150.975);

\path [draw=steelblue52138189, thick]
(axis cs:0.035801574587822,153.45)
--(axis cs:1.36053836345673,153.45);

\path [draw=steelblue52138189, ultra thick]
(axis cs:0.588363394141197,153.45)
--(axis cs:0.970039948821068,153.45);

\path [draw=steelblue52138189, thick]
(axis cs:0.14565797150135,155.925)
--(axis cs:1.35442590713501,155.925);

\path [draw=steelblue52138189, ultra thick]
(axis cs:0.579709142446518,155.925)
--(axis cs:0.943393751978874,155.925);

\path [draw=steelblue52138189, thick]
(axis cs:0.160626113414764,158.4)
--(axis cs:1.41303658485413,158.4);

\path [draw=steelblue52138189, ultra thick]
(axis cs:0.58771151304245,158.4)
--(axis cs:0.941280394792557,158.4);

\path [draw=steelblue52138189, thick]
(axis cs:0.144877731800079,160.875)
--(axis cs:1.37373340129852,160.875);

\path [draw=steelblue52138189, ultra thick]
(axis cs:0.602309122681618,160.875)
--(axis cs:0.960759490728378,160.875);

\path [draw=steelblue52138189, thick]
(axis cs:0.139238968491554,163.35)
--(axis cs:1.30640757083893,163.35);

\path [draw=steelblue52138189, ultra thick]
(axis cs:0.578379854559898,163.35)
--(axis cs:0.938424587249756,163.35);

\path [draw=steelblue52138189, thick]
(axis cs:0.216671332716942,165.825)
--(axis cs:1.45973646640778,165.825);

\path [draw=steelblue52138189, ultra thick]
(axis cs:0.561582148075104,165.825)
--(axis cs:0.929481476545334,165.825);

\path [draw=steelblue52138189, thick]
(axis cs:0.0395661517977715,168.3)
--(axis cs:1.40875971317291,168.3);

\path [draw=steelblue52138189, ultra thick]
(axis cs:0.562032163143158,168.3)
--(axis cs:0.975420191884041,168.3);

\addplot [thick, steelblue52138189, mark=*, mark size=2.25, mark options={solid,fill=whitesmoke238}, only marks]
table {%
0.791106909513474 0
};
\addplot [thick, steelblue52138189, mark=*, mark size=2.25, mark options={solid,fill=whitesmoke238}, only marks]
table {%
0.7855264544487 2.475
};
\addplot [thick, steelblue52138189, mark=*, mark size=2.25, mark options={solid,fill=whitesmoke238}, only marks]
table {%
0.742458343505859 4.95
};
\addplot [thick, steelblue52138189, mark=*, mark size=2.25, mark options={solid,fill=whitesmoke238}, only marks]
table {%
0.758943140506744 7.425
};
\addplot [thick, steelblue52138189, mark=*, mark size=2.25, mark options={solid,fill=whitesmoke238}, only marks]
table {%
0.790138840675354 9.9
};
\addplot [thick, steelblue52138189, mark=*, mark size=2.25, mark options={solid,fill=whitesmoke238}, only marks]
table {%
0.794582396745682 12.375
};
\addplot [thick, steelblue52138189, mark=*, mark size=2.25, mark options={solid,fill=whitesmoke238}, only marks]
table {%
0.777642279863358 14.85
};
\addplot [thick, steelblue52138189, mark=*, mark size=2.25, mark options={solid,fill=whitesmoke238}, only marks]
table {%
0.718494147062302 17.325
};
\addplot [thick, steelblue52138189, mark=*, mark size=2.25, mark options={solid,fill=whitesmoke238}, only marks]
table {%
0.774805307388306 19.8
};
\addplot [thick, steelblue52138189, mark=*, mark size=2.25, mark options={solid,fill=whitesmoke238}, only marks]
table {%
0.779260039329529 22.275
};
\addplot [thick, steelblue52138189, mark=*, mark size=2.25, mark options={solid,fill=whitesmoke238}, only marks]
table {%
0.789176732301712 24.75
};
\addplot [thick, steelblue52138189, mark=*, mark size=2.25, mark options={solid,fill=whitesmoke238}, only marks]
table {%
0.787952035665512 27.225
};
\addplot [thick, steelblue52138189, mark=*, mark size=2.25, mark options={solid,fill=whitesmoke238}, only marks]
table {%
0.785120278596878 29.7
};
\addplot [thick, steelblue52138189, mark=*, mark size=2.25, mark options={solid,fill=whitesmoke238}, only marks]
table {%
0.779612362384796 32.175
};
\addplot [thick, steelblue52138189, mark=*, mark size=2.25, mark options={solid,fill=whitesmoke238}, only marks]
table {%
0.79885596036911 34.65
};
\addplot [thick, steelblue52138189, mark=*, mark size=2.25, mark options={solid,fill=whitesmoke238}, only marks]
table {%
0.778644979000092 37.125
};
\addplot [thick, steelblue52138189, mark=*, mark size=2.25, mark options={solid,fill=whitesmoke238}, only marks]
table {%
0.780828505754471 39.6
};
\addplot [thick, steelblue52138189, mark=*, mark size=2.25, mark options={solid,fill=whitesmoke238}, only marks]
table {%
0.780847012996674 42.075
};
\addplot [thick, steelblue52138189, mark=*, mark size=2.25, mark options={solid,fill=whitesmoke238}, only marks]
table {%
0.710944294929504 44.55
};
\addplot [thick, steelblue52138189, mark=*, mark size=2.25, mark options={solid,fill=whitesmoke238}, only marks]
table {%
0.799208879470825 47.025
};
\addplot [thick, steelblue52138189, mark=*, mark size=2.25, mark options={solid,fill=whitesmoke238}, only marks]
table {%
0.797510147094727 49.5
};
\addplot [thick, steelblue52138189, mark=*, mark size=2.25, mark options={solid,fill=whitesmoke238}, only marks]
table {%
0.804191797971725 51.975
};
\addplot [thick, steelblue52138189, mark=*, mark size=2.25, mark options={solid,fill=whitesmoke238}, only marks]
table {%
0.773357599973679 54.45
};
\addplot [thick, steelblue52138189, mark=*, mark size=2.25, mark options={solid,fill=whitesmoke238}, only marks]
table {%
0.778060168027878 56.925
};
\addplot [thick, steelblue52138189, mark=*, mark size=2.25, mark options={solid,fill=whitesmoke238}, only marks]
table {%
0.741693526506424 59.4
};
\addplot [thick, steelblue52138189, mark=*, mark size=2.25, mark options={solid,fill=whitesmoke238}, only marks]
table {%
0.773068875074387 61.875
};
\addplot [thick, steelblue52138189, mark=*, mark size=2.25, mark options={solid,fill=whitesmoke238}, only marks]
table {%
0.782248258590698 64.35
};
\addplot [thick, steelblue52138189, mark=*, mark size=2.25, mark options={solid,fill=whitesmoke238}, only marks]
table {%
0.769261687994003 66.825
};
\addplot [thick, steelblue52138189, mark=*, mark size=2.25, mark options={solid,fill=whitesmoke238}, only marks]
table {%
0.769011110067368 69.3
};
\addplot [thick, steelblue52138189, mark=*, mark size=2.25, mark options={solid,fill=whitesmoke238}, only marks]
table {%
0.777936011552811 71.775
};
\addplot [thick, steelblue52138189, mark=*, mark size=2.25, mark options={solid,fill=whitesmoke238}, only marks]
table {%
0.85644456744194 74.25
};
\addplot [thick, steelblue52138189, mark=*, mark size=2.25, mark options={solid,fill=whitesmoke238}, only marks]
table {%
0.756614565849304 76.725
};
\addplot [thick, steelblue52138189, mark=*, mark size=2.25, mark options={solid,fill=whitesmoke238}, only marks]
table {%
0.78566786646843 79.2
};
\addplot [thick, steelblue52138189, mark=*, mark size=2.25, mark options={solid,fill=whitesmoke238}, only marks]
table {%
0.792114645242691 81.675
};
\addplot [thick, steelblue52138189, mark=*, mark size=2.25, mark options={solid,fill=whitesmoke238}, only marks]
table {%
0.776661604642868 84.15
};
\addplot [thick, steelblue52138189, mark=*, mark size=2.25, mark options={solid,fill=whitesmoke238}, only marks]
table {%
0.772822678089142 86.625
};
\addplot [thick, steelblue52138189, mark=*, mark size=2.25, mark options={solid,fill=whitesmoke238}, only marks]
table {%
0.819715976715088 89.1
};
\addplot [thick, steelblue52138189, mark=*, mark size=2.25, mark options={solid,fill=whitesmoke238}, only marks]
table {%
0.775874584913254 91.575
};
\addplot [thick, steelblue52138189, mark=*, mark size=2.25, mark options={solid,fill=whitesmoke238}, only marks]
table {%
0.793251514434814 94.05
};
\addplot [thick, steelblue52138189, mark=*, mark size=2.25, mark options={solid,fill=whitesmoke238}, only marks]
table {%
0.785898119211197 96.5249999999999
};
\addplot [thick, steelblue52138189, mark=*, mark size=2.25, mark options={solid,fill=whitesmoke238}, only marks]
table {%
0.77237993478775 98.9999999999999
};
\addplot [thick, steelblue52138189, mark=*, mark size=2.25, mark options={solid,fill=whitesmoke238}, only marks]
table {%
0.779096871614456 101.475
};
\addplot [thick, steelblue52138189, mark=*, mark size=2.25, mark options={solid,fill=whitesmoke238}, only marks]
table {%
0.728603512048721 103.95
};
\addplot [thick, steelblue52138189, mark=*, mark size=2.25, mark options={solid,fill=whitesmoke238}, only marks]
table {%
0.777023792266846 106.425
};
\addplot [thick, steelblue52138189, mark=*, mark size=2.25, mark options={solid,fill=whitesmoke238}, only marks]
table {%
0.770006209611893 108.9
};
\addplot [thick, steelblue52138189, mark=*, mark size=2.25, mark options={solid,fill=whitesmoke238}, only marks]
table {%
0.766152709722519 111.375
};
\addplot [thick, steelblue52138189, mark=*, mark size=2.25, mark options={solid,fill=whitesmoke238}, only marks]
table {%
0.772309154272079 113.85
};
\addplot [thick, steelblue52138189, mark=*, mark size=2.25, mark options={solid,fill=whitesmoke238}, only marks]
table {%
0.780234456062317 116.325
};
\addplot [thick, steelblue52138189, mark=*, mark size=2.25, mark options={solid,fill=whitesmoke238}, only marks]
table {%
0.793745100498199 118.8
};
\addplot [thick, steelblue52138189, mark=*, mark size=2.25, mark options={solid,fill=whitesmoke238}, only marks]
table {%
0.795571774244308 121.275
};
\addplot [thick, steelblue52138189, mark=*, mark size=2.25, mark options={solid,fill=whitesmoke238}, only marks]
table {%
0.756158977746964 123.75
};
\addplot [thick, steelblue52138189, mark=*, mark size=2.25, mark options={solid,fill=whitesmoke238}, only marks]
table {%
0.76534366607666 126.225
};
\addplot [thick, steelblue52138189, mark=*, mark size=2.25, mark options={solid,fill=whitesmoke238}, only marks]
table {%
0.762879073619843 128.7
};
\addplot [thick, steelblue52138189, mark=*, mark size=2.25, mark options={solid,fill=whitesmoke238}, only marks]
table {%
0.74093359708786 131.175
};
\addplot [thick, steelblue52138189, mark=*, mark size=2.25, mark options={solid,fill=whitesmoke238}, only marks]
table {%
0.786003321409225 133.65
};
\addplot [thick, steelblue52138189, mark=*, mark size=2.25, mark options={solid,fill=whitesmoke238}, only marks]
table {%
0.768572092056274 136.125
};
\addplot [thick, steelblue52138189, mark=*, mark size=2.25, mark options={solid,fill=whitesmoke238}, only marks]
table {%
0.77736759185791 138.6
};
\addplot [thick, steelblue52138189, mark=*, mark size=2.25, mark options={solid,fill=whitesmoke238}, only marks]
table {%
0.812776148319244 141.075
};
\addplot [thick, steelblue52138189, mark=*, mark size=2.25, mark options={solid,fill=whitesmoke238}, only marks]
table {%
0.794874429702759 143.55
};
\addplot [thick, steelblue52138189, mark=*, mark size=2.25, mark options={solid,fill=whitesmoke238}, only marks]
table {%
0.765951484441757 146.025
};
\addplot [thick, steelblue52138189, mark=*, mark size=2.25, mark options={solid,fill=whitesmoke238}, only marks]
table {%
0.784937262535095 148.5
};
\addplot [thick, steelblue52138189, mark=*, mark size=2.25, mark options={solid,fill=whitesmoke238}, only marks]
table {%
0.780945926904678 150.975
};
\addplot [thick, steelblue52138189, mark=*, mark size=2.25, mark options={solid,fill=whitesmoke238}, only marks]
table {%
0.780886620283127 153.45
};
\addplot [thick, steelblue52138189, mark=*, mark size=2.25, mark options={solid,fill=whitesmoke238}, only marks]
table {%
0.751143008470535 155.925
};
\addplot [thick, steelblue52138189, mark=*, mark size=2.25, mark options={solid,fill=whitesmoke238}, only marks]
table {%
0.778606414794922 158.4
};
\addplot [thick, steelblue52138189, mark=*, mark size=2.25, mark options={solid,fill=whitesmoke238}, only marks]
table {%
0.792304039001465 160.875
};
\addplot [thick, steelblue52138189, mark=*, mark size=2.25, mark options={solid,fill=whitesmoke238}, only marks]
table {%
0.763363510370255 163.35
};
\addplot [thick, steelblue52138189, mark=*, mark size=2.25, mark options={solid,fill=whitesmoke238}, only marks]
table {%
0.750214725732803 165.825
};
\addplot [thick, steelblue52138189, mark=*, mark size=2.25, mark options={solid,fill=whitesmoke238}, only marks]
table {%
0.787744760513306 168.3
};
\end{axis}

\end{tikzpicture}

%% file: figures/student-bias.tex
\begin{tikzpicture}

\definecolor{darkgray178}{RGB}{178,178,178}
\definecolor{silver188}{RGB}{188,188,188}
\definecolor{steelblue52138189}{RGB}{52,138,189}
\definecolor{whitesmoke238}{RGB}{238,238,238}

\begin{axis}[
axis background/.style={fill=whitesmoke238},
axis line style={silver188},
title={Student Bias Distributions},
width=\columnwidth,
height=\columnwidth,
x grid style={darkgray178},
xmajorgrids,
xmin=-3.59335938692093, xmax=3.74517885446548,
xtick pos=left,
xtick style={color=black},
y grid style={darkgray178},
ymajorticks=false,
ymin=-1.35, ymax=171,
ytick style={color=black},
ytick={0,2.475,4.95,7.425,9.9,12.375,14.85,17.325,19.8,22.275,24.75,27.225,29.7,32.175,34.65,37.125,39.6,42.075,44.55,47.025,49.5,51.975,54.45,56.925,59.4,61.875,64.35,66.825,69.3,71.775,74.25,76.725,79.2,81.675,84.15,86.625,89.1,91.575,94.05,96.5249999999999,98.9999999999999,101.475,103.95,106.425,108.9,111.375,113.85,116.325,118.8,121.275,123.75,126.225,128.7,131.175,133.65,136.125,138.6,141.075,143.55,146.025,148.5,150.975,153.45,155.925,158.4,160.875,163.35,165.825,168.3},
yticklabels={
  [68],
  [67],
  [66],
  [65],
  [64],
  [63],
  [62],
  [61],
  [60],
  [59],
  [58],
  [57],
  [56],
  [55],
  [54],
  [53],
  [52],
  [51],
  [50],
  [49],
  [48],
  [47],
  [46],
  [45],
  [44],
  [43],
  [42],
  [41],
  [40],
  [39],
  [38],
  [37],
  [36],
  [35],
  [34],
  [33],
  [32],
  [31],
  [30],
  [29],
  [28],
  [27],
  [26],
  [25],
  [24],
  [23],
  [22],
  [21],
  [20],
  [19],
  [18],
  [17],
  [16],
  [15],
  [14],
  [13],
  [12],
  [11],
  [10],
  [9],
  [8],
  [7],
  [6],
  [5],
  [4],
  [3],
  [2],
  [1],
  β[0]
}
]
\path [draw=steelblue52138189, thick]
(axis cs:-1.17642033100128,0)
--(axis cs:2.51585245132446,0);

\path [draw=steelblue52138189, ultra thick]
(axis cs:-0.0356865217909217,0)
--(axis cs:1.19687378406525,0);

\path [draw=steelblue52138189, thick]
(axis cs:-1.60119962692261,2.475)
--(axis cs:2.1646831035614,2.475);

\path [draw=steelblue52138189, ultra thick]
(axis cs:-0.29542575776577,2.475)
--(axis cs:0.967361330986023,2.475);

\path [draw=steelblue52138189, thick]
(axis cs:-1.26158475875854,4.95)
--(axis cs:2.48757076263428,4.95);

\path [draw=steelblue52138189, ultra thick]
(axis cs:0.238764341920614,4.95)
--(axis cs:1.3821342587471,4.95);

\path [draw=steelblue52138189, thick]
(axis cs:-1.53359055519104,7.425)
--(axis cs:1.96287798881531,7.425);

\path [draw=steelblue52138189, ultra thick]
(axis cs:-0.324393510818481,7.425)
--(axis cs:0.937260851264,7.425);

\path [draw=steelblue52138189, thick]
(axis cs:-1.65186762809753,9.9)
--(axis cs:2.21196985244751,9.9);

\path [draw=steelblue52138189, ultra thick]
(axis cs:-0.315438911318779,9.9)
--(axis cs:0.993529781699181,9.9);

\path [draw=steelblue52138189, thick]
(axis cs:-0.930849432945251,12.375)
--(axis cs:2.57910442352295,12.375);

\path [draw=steelblue52138189, ultra thick]
(axis cs:0.345466837286949,12.375)
--(axis cs:1.48997312784195,12.375);

\path [draw=steelblue52138189, thick]
(axis cs:-0.718868911266327,14.85)
--(axis cs:2.8289647102356,14.85);

\path [draw=steelblue52138189, ultra thick]
(axis cs:0.493937104940414,14.85)
--(axis cs:1.77710542082787,14.85);

\path [draw=steelblue52138189, thick]
(axis cs:-1.86322462558746,17.325)
--(axis cs:1.9015200138092,17.325);

\path [draw=steelblue52138189, ultra thick]
(axis cs:-0.656170606613159,17.325)
--(axis cs:0.64874230325222,17.325);

\path [draw=steelblue52138189, thick]
(axis cs:-2.47541332244873,19.8)
--(axis cs:1.23567473888397,19.8);

\path [draw=steelblue52138189, ultra thick]
(axis cs:-1.05941817164421,19.8)
--(axis cs:0.0570716839283705,19.8);

\path [draw=steelblue52138189, thick]
(axis cs:-1.31294786930084,22.275)
--(axis cs:2.36803793907166,22.275);

\path [draw=steelblue52138189, ultra thick]
(axis cs:-0.0863414984196424,22.275)
--(axis cs:1.05866098403931,22.275);

\path [draw=steelblue52138189, thick]
(axis cs:-1.73497593402863,24.75)
--(axis cs:1.85862684249878,24.75);

\path [draw=steelblue52138189, ultra thick]
(axis cs:-0.506611943244934,24.75)
--(axis cs:0.773548498749733,24.75);

\path [draw=steelblue52138189, thick]
(axis cs:-1.42858803272247,27.225)
--(axis cs:2.07734155654907,27.225);

\path [draw=steelblue52138189, ultra thick]
(axis cs:0.0195983992889524,27.225)
--(axis cs:1.09320840239525,27.225);

\path [draw=steelblue52138189, thick]
(axis cs:-1.16588175296783,29.7)
--(axis cs:2.31789374351501,29.7);

\path [draw=steelblue52138189, ultra thick]
(axis cs:-0.0145745640620589,29.7)
--(axis cs:1.15603217482567,29.7);

\path [draw=steelblue52138189, thick]
(axis cs:-0.907051563262939,32.175)
--(axis cs:2.91773247718811,32.175);

\path [draw=steelblue52138189, ultra thick]
(axis cs:0.542745977640152,32.175)
--(axis cs:1.69767141342163,32.175);

\path [draw=steelblue52138189, thick]
(axis cs:-0.774311184883118,34.65)
--(axis cs:2.95460653305054,34.65);

\path [draw=steelblue52138189, ultra thick]
(axis cs:0.236228760331869,34.65)
--(axis cs:1.42185744643211,34.65);

\path [draw=steelblue52138189, thick]
(axis cs:-1.50752544403076,37.125)
--(axis cs:2.02401900291443,37.125);

\path [draw=steelblue52138189, ultra thick]
(axis cs:-0.112321274355054,37.125)
--(axis cs:1.09556198120117,37.125);

\path [draw=steelblue52138189, thick]
(axis cs:-2.63584876060486,39.6)
--(axis cs:0.8351891040802,39.6);

\path [draw=steelblue52138189, ultra thick]
(axis cs:-1.27913051843643,39.6)
--(axis cs:-0.229143612086773,39.6);

\path [draw=steelblue52138189, thick]
(axis cs:-1.42874360084534,42.075)
--(axis cs:2.15183687210083,42.075);

\path [draw=steelblue52138189, ultra thick]
(axis cs:-0.370314761996269,42.075)
--(axis cs:0.885003596544266,42.075);

\path [draw=steelblue52138189, thick]
(axis cs:-2.50851154327393,44.55)
--(axis cs:1.07741189002991,44.55);

\path [draw=steelblue52138189, ultra thick]
(axis cs:-1.22566124796867,44.55)
--(axis cs:-0.0596413156017661,44.55);

\path [draw=steelblue52138189, thick]
(axis cs:-1.18163847923279,47.025)
--(axis cs:2.37644529342651,47.025);

\path [draw=steelblue52138189, ultra thick]
(axis cs:0.27168395370245,47.025)
--(axis cs:1.44671601057053,47.025);

\path [draw=steelblue52138189, thick]
(axis cs:-2.09480547904968,49.5)
--(axis cs:1.56314659118652,49.5);

\path [draw=steelblue52138189, ultra thick]
(axis cs:-0.747556522488594,49.5)
--(axis cs:0.520437493920326,49.5);

\path [draw=steelblue52138189, thick]
(axis cs:-2.79148769378662,51.975)
--(axis cs:0.890205144882202,51.975);

\path [draw=steelblue52138189, ultra thick]
(axis cs:-1.63355284929276,51.975)
--(axis cs:-0.392009392380714,51.975);

\path [draw=steelblue52138189, thick]
(axis cs:-0.872154295444489,54.45)
--(axis cs:2.55660772323608,54.45);

\path [draw=steelblue52138189, ultra thick]
(axis cs:0.120157448574901,54.45)
--(axis cs:1.29218220710754,54.45);

\path [draw=steelblue52138189, thick]
(axis cs:-0.327756732702255,56.925)
--(axis cs:3.21523094177246,56.925);

\path [draw=steelblue52138189, ultra thick]
(axis cs:0.913301467895508,56.925)
--(axis cs:2.03340381383896,56.925);

\path [draw=steelblue52138189, thick]
(axis cs:-1.00864088535309,59.4)
--(axis cs:2.59371423721313,59.4);

\path [draw=steelblue52138189, ultra thick]
(axis cs:0.153012599796057,59.4)
--(axis cs:1.35937008261681,59.4);

\path [draw=steelblue52138189, thick]
(axis cs:-1.6585156917572,61.875)
--(axis cs:2.11491441726685,61.875);

\path [draw=steelblue52138189, ultra thick]
(axis cs:-0.39030596613884,61.875)
--(axis cs:0.801572367548943,61.875);

\path [draw=steelblue52138189, thick]
(axis cs:-0.595535159111023,64.35)
--(axis cs:2.83950972557068,64.35);

\path [draw=steelblue52138189, ultra thick]
(axis cs:0.726824104785919,64.35)
--(axis cs:1.78220742940903,64.35);

\path [draw=steelblue52138189, thick]
(axis cs:-1.53204655647278,66.825)
--(axis cs:2.10918807983398,66.825);

\path [draw=steelblue52138189, ultra thick]
(axis cs:-0.291894160211086,66.825)
--(axis cs:0.983855545520782,66.825);

\path [draw=steelblue52138189, thick]
(axis cs:-0.772355377674103,69.3)
--(axis cs:2.79206681251526,69.3);

\path [draw=steelblue52138189, ultra thick]
(axis cs:0.523647502064705,69.3)
--(axis cs:1.81698513031006,69.3);

\path [draw=steelblue52138189, thick]
(axis cs:-0.821269631385803,71.775)
--(axis cs:2.72111773490906,71.775);

\path [draw=steelblue52138189, ultra thick]
(axis cs:0.489580079913139,71.775)
--(axis cs:1.71751245856285,71.775);

\path [draw=steelblue52138189, thick]
(axis cs:-3.25427389144897,74.25)
--(axis cs:0.400959432125092,74.25);

\path [draw=steelblue52138189, ultra thick]
(axis cs:-2.2082827091217,74.25)
--(axis cs:-0.864758118987083,74.25);

\path [draw=steelblue52138189, thick]
(axis cs:-0.121896132826805,76.725)
--(axis cs:3.41160893440247,76.725);

\path [draw=steelblue52138189, ultra thick]
(axis cs:0.9493328332901,76.725)
--(axis cs:2.069071829319,76.725);

\path [draw=steelblue52138189, thick]
(axis cs:-0.437569469213486,79.2)
--(axis cs:3.05972456932068,79.2);

\path [draw=steelblue52138189, ultra thick]
(axis cs:0.671073362231255,79.2)
--(axis cs:1.78677809238434,79.2);

\path [draw=steelblue52138189, thick]
(axis cs:-1.05204129219055,81.675)
--(axis cs:2.60034489631653,81.675);

\path [draw=steelblue52138189, ultra thick]
(axis cs:0.339378796517849,81.675)
--(axis cs:1.57466530799866,81.675);

\path [draw=steelblue52138189, thick]
(axis cs:-0.672797322273254,84.15)
--(axis cs:2.86809134483337,84.15);

\path [draw=steelblue52138189, ultra thick]
(axis cs:0.473187789320946,84.15)
--(axis cs:1.66458439826965,84.15);

\path [draw=steelblue52138189, thick]
(axis cs:-0.821691930294037,86.625)
--(axis cs:2.81031799316406,86.625);

\path [draw=steelblue52138189, ultra thick]
(axis cs:0.447526931762695,86.625)
--(axis cs:1.63814169168472,86.625);

\path [draw=steelblue52138189, thick]
(axis cs:-1.16212844848633,89.1)
--(axis cs:2.3437077999115,89.1);

\path [draw=steelblue52138189, ultra thick]
(axis cs:-0.0499665318056941,89.1)
--(axis cs:1.21179232001305,89.1);

\path [draw=steelblue52138189, thick]
(axis cs:-1.33800160884857,91.575)
--(axis cs:2.38918352127075,91.575);

\path [draw=steelblue52138189, ultra thick]
(axis cs:-0.36377327144146,91.575)
--(axis cs:0.995824038982391,91.575);

\path [draw=steelblue52138189, thick]
(axis cs:-1.53040528297424,94.05)
--(axis cs:2.28915524482727,94.05);

\path [draw=steelblue52138189, ultra thick]
(axis cs:-0.242573969066143,94.05)
--(axis cs:0.998799130320549,94.05);

\path [draw=steelblue52138189, thick]
(axis cs:-1.03281056880951,96.5249999999999)
--(axis cs:2.4499077796936,96.5249999999999);

\path [draw=steelblue52138189, ultra thick]
(axis cs:0.206759661436081,96.5249999999999)
--(axis cs:1.38448819518089,96.5249999999999);

\path [draw=steelblue52138189, thick]
(axis cs:-1.33521854877472,98.9999999999999)
--(axis cs:2.45682835578918,98.9999999999999);

\path [draw=steelblue52138189, ultra thick]
(axis cs:-0.132786765694618,98.9999999999999)
--(axis cs:1.09112185239792,98.9999999999999);

\path [draw=steelblue52138189, thick]
(axis cs:-0.15042170882225,101.475)
--(axis cs:3.29797387123108,101.475);

\path [draw=steelblue52138189, ultra thick]
(axis cs:0.903315022587776,101.475)
--(axis cs:2.04964107275009,101.475);

\path [draw=steelblue52138189, thick]
(axis cs:-0.786851942539215,103.95)
--(axis cs:2.96218395233154,103.95);

\path [draw=steelblue52138189, ultra thick]
(axis cs:0.53719362616539,103.95)
--(axis cs:1.73960748314857,103.95);

\path [draw=steelblue52138189, thick]
(axis cs:-1.05751013755798,106.425)
--(axis cs:2.38007140159607,106.425);

\path [draw=steelblue52138189, ultra thick]
(axis cs:0.0952788088470697,106.425)
--(axis cs:1.24059844017029,106.425);

\path [draw=steelblue52138189, thick]
(axis cs:-1.4863988161087,108.9)
--(axis cs:1.96147954463959,108.9);

\path [draw=steelblue52138189, ultra thick]
(axis cs:-0.213012795895338,108.9)
--(axis cs:0.985563993453979,108.9);

\path [draw=steelblue52138189, thick]
(axis cs:-0.94072562456131,111.375)
--(axis cs:2.8381507396698,111.375);

\path [draw=steelblue52138189, ultra thick]
(axis cs:0.197826985269785,111.375)
--(axis cs:1.50449511408806,111.375);

\path [draw=steelblue52138189, thick]
(axis cs:-1.71776270866394,113.85)
--(axis cs:2.24506568908691,113.85);

\path [draw=steelblue52138189, ultra thick]
(axis cs:-0.120221359655261,113.85)
--(axis cs:1.10904693603516,113.85);

\path [draw=steelblue52138189, thick]
(axis cs:-1.8664870262146,116.325)
--(axis cs:1.95690274238586,116.325);

\path [draw=steelblue52138189, ultra thick]
(axis cs:-0.400194890797138,116.325)
--(axis cs:0.813872367143631,116.325);

\path [draw=steelblue52138189, thick]
(axis cs:-1.29570770263672,118.8)
--(axis cs:2.55317854881287,118.8);

\path [draw=steelblue52138189, ultra thick]
(axis cs:-0.327894575893879,118.8)
--(axis cs:0.998576030135155,118.8);

\path [draw=steelblue52138189, thick]
(axis cs:-1.11960828304291,121.275)
--(axis cs:2.47812676429749,121.275);

\path [draw=steelblue52138189, ultra thick]
(axis cs:0.0151020311750472,121.275)
--(axis cs:1.14273056387901,121.275);

\path [draw=steelblue52138189, thick]
(axis cs:-1.34971177577972,123.75)
--(axis cs:2.23383975028992,123.75);

\path [draw=steelblue52138189, ultra thick]
(axis cs:-0.26916491985321,123.75)
--(axis cs:0.965137362480164,123.75);

\path [draw=steelblue52138189, thick]
(axis cs:-1.18065404891968,126.225)
--(axis cs:2.59637522697449,126.225);

\path [draw=steelblue52138189, ultra thick]
(axis cs:0.0147369727492332,126.225)
--(axis cs:1.26385432481766,126.225);

\path [draw=steelblue52138189, thick]
(axis cs:-1.21706604957581,128.7)
--(axis cs:2.28641033172607,128.7);

\path [draw=steelblue52138189, ultra thick]
(axis cs:-0.0497427126392722,128.7)
--(axis cs:1.10836723446846,128.7);

\path [draw=steelblue52138189, thick]
(axis cs:-1.07776939868927,131.175)
--(axis cs:2.35825204849243,131.175);

\path [draw=steelblue52138189, ultra thick]
(axis cs:-0.203089714050293,131.175)
--(axis cs:1.17742672562599,131.175);

\path [draw=steelblue52138189, thick]
(axis cs:-1.22611129283905,133.65)
--(axis cs:2.54360604286194,133.65);

\path [draw=steelblue52138189, ultra thick]
(axis cs:-0.00844860286451876,133.65)
--(axis cs:1.17922991514206,133.65);

\path [draw=steelblue52138189, thick]
(axis cs:-1.11139214038849,136.125)
--(axis cs:2.32477760314941,136.125);

\path [draw=steelblue52138189, ultra thick]
(axis cs:0.0677907261997461,136.125)
--(axis cs:1.26472821831703,136.125);

\path [draw=steelblue52138189, thick]
(axis cs:-0.865119636058807,138.6)
--(axis cs:2.65097498893738,138.6);

\path [draw=steelblue52138189, ultra thick]
(axis cs:0.120443737134337,138.6)
--(axis cs:1.32431742548943,138.6);

\path [draw=steelblue52138189, thick]
(axis cs:-3.25978946685791,141.075)
--(axis cs:0.30904832482338,141.075);

\path [draw=steelblue52138189, ultra thick]
(axis cs:-2.15588307380676,141.075)
--(axis cs:-0.936952739953995,141.075);

\path [draw=steelblue52138189, thick]
(axis cs:-0.875838696956635,143.55)
--(axis cs:2.76753163337708,143.55);

\path [draw=steelblue52138189, ultra thick]
(axis cs:0.228541858494282,143.55)
--(axis cs:1.44343730807304,143.55);

\path [draw=steelblue52138189, thick]
(axis cs:-1.26418232917786,146.025)
--(axis cs:2.40538454055786,146.025);

\path [draw=steelblue52138189, ultra thick]
(axis cs:-0.0903485249727964,146.025)
--(axis cs:1.077566832304,146.025);

\path [draw=steelblue52138189, thick]
(axis cs:-1.35958325862885,148.5)
--(axis cs:2.38371253013611,148.5);

\path [draw=steelblue52138189, ultra thick]
(axis cs:-0.189157050102949,148.5)
--(axis cs:1.14484629034996,148.5);

\path [draw=steelblue52138189, thick]
(axis cs:-0.982549011707306,150.975)
--(axis cs:2.79659748077393,150.975);

\path [draw=steelblue52138189, ultra thick]
(axis cs:0.141453444957733,150.975)
--(axis cs:1.41080868244171,150.975);

\path [draw=steelblue52138189, thick]
(axis cs:-1.07401657104492,153.45)
--(axis cs:2.67077970504761,153.45);

\path [draw=steelblue52138189, ultra thick]
(axis cs:0.191015623509884,153.45)
--(axis cs:1.41652452945709,153.45);

\path [draw=steelblue52138189, thick]
(axis cs:-1.4842723608017,155.925)
--(axis cs:2.21413421630859,155.925);

\path [draw=steelblue52138189, ultra thick]
(axis cs:-0.309116646647453,155.925)
--(axis cs:0.892242357134819,155.925);

\path [draw=steelblue52138189, thick]
(axis cs:-1.85467278957367,158.4)
--(axis cs:1.73161840438843,158.4);

\path [draw=steelblue52138189, ultra thick]
(axis cs:-0.652401506900787,158.4)
--(axis cs:0.535633832216263,158.4);

\path [draw=steelblue52138189, thick]
(axis cs:-1.29254603385925,160.875)
--(axis cs:2.25010418891907,160.875);

\path [draw=steelblue52138189, ultra thick]
(axis cs:-0.182274755090475,160.875)
--(axis cs:0.98781755566597,160.875);

\path [draw=steelblue52138189, thick]
(axis cs:-1.48914706707001,163.35)
--(axis cs:2.17329430580139,163.35);

\path [draw=steelblue52138189, ultra thick]
(axis cs:-0.151605978608131,163.35)
--(axis cs:1.01416218280792,163.35);

\path [draw=steelblue52138189, thick]
(axis cs:-1.43737542629242,165.825)
--(axis cs:2.05480265617371,165.825);

\path [draw=steelblue52138189, ultra thick]
(axis cs:-0.168875727802515,165.825)
--(axis cs:0.998740509152412,165.825);

\path [draw=steelblue52138189, thick]
(axis cs:-0.49628672003746,168.3)
--(axis cs:2.87328386306763,168.3);

\path [draw=steelblue52138189, ultra thick]
(axis cs:0.569034278392792,168.3)
--(axis cs:1.76784920692444,168.3);

\addplot [thick, steelblue52138189, mark=*, mark size=2.25, mark options={solid,fill=whitesmoke238}, only marks]
table {%
0.56122288107872 0
};
\addplot [thick, steelblue52138189, mark=*, mark size=2.25, mark options={solid,fill=whitesmoke238}, only marks]
table {%
0.294727697968483 2.475
};
\addplot [thick, steelblue52138189, mark=*, mark size=2.25, mark options={solid,fill=whitesmoke238}, only marks]
table {%
0.81560081243515 4.95
};
\addplot [thick, steelblue52138189, mark=*, mark size=2.25, mark options={solid,fill=whitesmoke238}, only marks]
table {%
0.241971477866173 7.425
};
\addplot [thick, steelblue52138189, mark=*, mark size=2.25, mark options={solid,fill=whitesmoke238}, only marks]
table {%
0.313420683145523 9.9
};
\addplot [thick, steelblue52138189, mark=*, mark size=2.25, mark options={solid,fill=whitesmoke238}, only marks]
table {%
0.887606292963028 12.375
};
\addplot [thick, steelblue52138189, mark=*, mark size=2.25, mark options={solid,fill=whitesmoke238}, only marks]
table {%
1.09787219762802 14.85
};
\addplot [thick, steelblue52138189, mark=*, mark size=2.25, mark options={solid,fill=whitesmoke238}, only marks]
table {%
-0.0461887177079916 17.325
};
\addplot [thick, steelblue52138189, mark=*, mark size=2.25, mark options={solid,fill=whitesmoke238}, only marks]
table {%
-0.500270336866379 19.8
};
\addplot [thick, steelblue52138189, mark=*, mark size=2.25, mark options={solid,fill=whitesmoke238}, only marks]
table {%
0.453608989715576 22.275
};
\addplot [thick, steelblue52138189, mark=*, mark size=2.25, mark options={solid,fill=whitesmoke238}, only marks]
table {%
0.126492951065302 24.75
};
\addplot [thick, steelblue52138189, mark=*, mark size=2.25, mark options={solid,fill=whitesmoke238}, only marks]
table {%
0.494995698332787 27.225
};
\addplot [thick, steelblue52138189, mark=*, mark size=2.25, mark options={solid,fill=whitesmoke238}, only marks]
table {%
0.557705014944077 29.7
};
\addplot [thick, steelblue52138189, mark=*, mark size=2.25, mark options={solid,fill=whitesmoke238}, only marks]
table {%
1.08040565252304 32.175
};
\addplot [thick, steelblue52138189, mark=*, mark size=2.25, mark options={solid,fill=whitesmoke238}, only marks]
table {%
0.849770218133926 34.65
};
\addplot [thick, steelblue52138189, mark=*, mark size=2.25, mark options={solid,fill=whitesmoke238}, only marks]
table {%
0.481618329882622 37.125
};
\addplot [thick, steelblue52138189, mark=*, mark size=2.25, mark options={solid,fill=whitesmoke238}, only marks]
table {%
-0.782447874546051 39.6
};
\addplot [thick, steelblue52138189, mark=*, mark size=2.25, mark options={solid,fill=whitesmoke238}, only marks]
table {%
0.217369392514229 42.075
};
\addplot [thick, steelblue52138189, mark=*, mark size=2.25, mark options={solid,fill=whitesmoke238}, only marks]
table {%
-0.636416018009186 44.55
};
\addplot [thick, steelblue52138189, mark=*, mark size=2.25, mark options={solid,fill=whitesmoke238}, only marks]
table {%
0.851349323987961 47.025
};
\addplot [thick, steelblue52138189, mark=*, mark size=2.25, mark options={solid,fill=whitesmoke238}, only marks]
table {%
-0.0871490947902203 49.5
};
\addplot [thick, steelblue52138189, mark=*, mark size=2.25, mark options={solid,fill=whitesmoke238}, only marks]
table {%
-1.0604088306427 51.975
};
\addplot [thick, steelblue52138189, mark=*, mark size=2.25, mark options={solid,fill=whitesmoke238}, only marks]
table {%
0.707348614931107 54.45
};
\addplot [thick, steelblue52138189, mark=*, mark size=2.25, mark options={solid,fill=whitesmoke238}, only marks]
table {%
1.46742755174637 56.925
};
\addplot [thick, steelblue52138189, mark=*, mark size=2.25, mark options={solid,fill=whitesmoke238}, only marks]
table {%
0.753226101398468 59.4
};
\addplot [thick, steelblue52138189, mark=*, mark size=2.25, mark options={solid,fill=whitesmoke238}, only marks]
table {%
0.210688307881355 61.875
};
\addplot [thick, steelblue52138189, mark=*, mark size=2.25, mark options={solid,fill=whitesmoke238}, only marks]
table {%
1.21181178092957 64.35
};
\addplot [thick, steelblue52138189, mark=*, mark size=2.25, mark options={solid,fill=whitesmoke238}, only marks]
table {%
0.318888664245605 66.825
};
\addplot [thick, steelblue52138189, mark=*, mark size=2.25, mark options={solid,fill=whitesmoke238}, only marks]
table {%
1.13204967975616 69.3
};
\addplot [thick, steelblue52138189, mark=*, mark size=2.25, mark options={solid,fill=whitesmoke238}, only marks]
table {%
1.08691650629044 71.775
};
\addplot [thick, steelblue52138189, mark=*, mark size=2.25, mark options={solid,fill=whitesmoke238}, only marks]
table {%
-1.52887916564941 74.25
};
\addplot [thick, steelblue52138189, mark=*, mark size=2.25, mark options={solid,fill=whitesmoke238}, only marks]
table {%
1.512830555439 76.725
};
\addplot [thick, steelblue52138189, mark=*, mark size=2.25, mark options={solid,fill=whitesmoke238}, only marks]
table {%
1.22144013643265 79.2
};
\addplot [thick, steelblue52138189, mark=*, mark size=2.25, mark options={solid,fill=whitesmoke238}, only marks]
table {%
0.94099086523056 81.675
};
\addplot [thick, steelblue52138189, mark=*, mark size=2.25, mark options={solid,fill=whitesmoke238}, only marks]
table {%
1.08896297216415 84.15
};
\addplot [thick, steelblue52138189, mark=*, mark size=2.25, mark options={solid,fill=whitesmoke238}, only marks]
table {%
1.018574655056 86.625
};
\addplot [thick, steelblue52138189, mark=*, mark size=2.25, mark options={solid,fill=whitesmoke238}, only marks]
table {%
0.580564111471176 89.1
};
\addplot [thick, steelblue52138189, mark=*, mark size=2.25, mark options={solid,fill=whitesmoke238}, only marks]
table {%
0.343698784708977 91.575
};
\addplot [thick, steelblue52138189, mark=*, mark size=2.25, mark options={solid,fill=whitesmoke238}, only marks]
table {%
0.379890143871307 94.05
};
\addplot [thick, steelblue52138189, mark=*, mark size=2.25, mark options={solid,fill=whitesmoke238}, only marks]
table {%
0.765103757381439 96.5249999999999
};
\addplot [thick, steelblue52138189, mark=*, mark size=2.25, mark options={solid,fill=whitesmoke238}, only marks]
table {%
0.455850005149841 98.9999999999999
};
\addplot [thick, steelblue52138189, mark=*, mark size=2.25, mark options={solid,fill=whitesmoke238}, only marks]
table {%
1.43167901039124 101.475
};
\addplot [thick, steelblue52138189, mark=*, mark size=2.25, mark options={solid,fill=whitesmoke238}, only marks]
table {%
1.08246898651123 103.95
};
\addplot [thick, steelblue52138189, mark=*, mark size=2.25, mark options={solid,fill=whitesmoke238}, only marks]
table {%
0.660625845193863 106.425
};
\addplot [thick, steelblue52138189, mark=*, mark size=2.25, mark options={solid,fill=whitesmoke238}, only marks]
table {%
0.388576120138168 108.9
};
\addplot [thick, steelblue52138189, mark=*, mark size=2.25, mark options={solid,fill=whitesmoke238}, only marks]
table {%
0.809033870697021 111.375
};
\addplot [thick, steelblue52138189, mark=*, mark size=2.25, mark options={solid,fill=whitesmoke238}, only marks]
table {%
0.531569391489029 113.85
};
\addplot [thick, steelblue52138189, mark=*, mark size=2.25, mark options={solid,fill=whitesmoke238}, only marks]
table {%
0.19047774374485 116.325
};
\addplot [thick, steelblue52138189, mark=*, mark size=2.25, mark options={solid,fill=whitesmoke238}, only marks]
table {%
0.334324181079865 118.8
};
\addplot [thick, steelblue52138189, mark=*, mark size=2.25, mark options={solid,fill=whitesmoke238}, only marks]
table {%
0.556954652070999 121.275
};
\addplot [thick, steelblue52138189, mark=*, mark size=2.25, mark options={solid,fill=whitesmoke238}, only marks]
table {%
0.352239817380905 123.75
};
\addplot [thick, steelblue52138189, mark=*, mark size=2.25, mark options={solid,fill=whitesmoke238}, only marks]
table {%
0.646183401346207 126.225
};
\addplot [thick, steelblue52138189, mark=*, mark size=2.25, mark options={solid,fill=whitesmoke238}, only marks]
table {%
0.540297329425812 128.7
};
\addplot [thick, steelblue52138189, mark=*, mark size=2.25, mark options={solid,fill=whitesmoke238}, only marks]
table {%
0.51353332400322 131.175
};
\addplot [thick, steelblue52138189, mark=*, mark size=2.25, mark options={solid,fill=whitesmoke238}, only marks]
table {%
0.640751719474792 133.65
};
\addplot [thick, steelblue52138189, mark=*, mark size=2.25, mark options={solid,fill=whitesmoke238}, only marks]
table {%
0.639436423778534 136.125
};
\addplot [thick, steelblue52138189, mark=*, mark size=2.25, mark options={solid,fill=whitesmoke238}, only marks]
table {%
0.75604373216629 138.6
};
\addplot [thick, steelblue52138189, mark=*, mark size=2.25, mark options={solid,fill=whitesmoke238}, only marks]
table {%
-1.51612901687622 141.075
};
\addplot [thick, steelblue52138189, mark=*, mark size=2.25, mark options={solid,fill=whitesmoke238}, only marks]
table {%
0.829534471035004 143.55
};
\addplot [thick, steelblue52138189, mark=*, mark size=2.25, mark options={solid,fill=whitesmoke238}, only marks]
table {%
0.546390891075134 146.025
};
\addplot [thick, steelblue52138189, mark=*, mark size=2.25, mark options={solid,fill=whitesmoke238}, only marks]
table {%
0.497412100434303 148.5
};
\addplot [thick, steelblue52138189, mark=*, mark size=2.25, mark options={solid,fill=whitesmoke238}, only marks]
table {%
0.747980058193207 150.975
};
\addplot [thick, steelblue52138189, mark=*, mark size=2.25, mark options={solid,fill=whitesmoke238}, only marks]
table {%
0.818281501531601 153.45
};
\addplot [thick, steelblue52138189, mark=*, mark size=2.25, mark options={solid,fill=whitesmoke238}, only marks]
table {%
0.273716241121292 155.925
};
\addplot [thick, steelblue52138189, mark=*, mark size=2.25, mark options={solid,fill=whitesmoke238}, only marks]
table {%
-0.102129150182009 158.4
};
\addplot [thick, steelblue52138189, mark=*, mark size=2.25, mark options={solid,fill=whitesmoke238}, only marks]
table {%
0.415027767419815 160.875
};
\addplot [thick, steelblue52138189, mark=*, mark size=2.25, mark options={solid,fill=whitesmoke238}, only marks]
table {%
0.418118074536324 163.35
};
\addplot [thick, steelblue52138189, mark=*, mark size=2.25, mark options={solid,fill=whitesmoke238}, only marks]
table {%
0.418342664837837 165.825
};
\addplot [thick, steelblue52138189, mark=*, mark size=2.25, mark options={solid,fill=whitesmoke238}, only marks]
table {%
1.15077787637711 168.3
};
\end{axis}

\end{tikzpicture}